\documentclass[aps,pra,reprint,amsmath,superscriptaddress,amssymb,bibnotes]{revtex4-1}

\usepackage[ansinew]{inputenc}
\usepackage[T1]{fontenc} 

\usepackage{graphicx}
\usepackage[colorlinks,linkcolor=black,citecolor=black,urlcolor=black]{hyperref}
\usepackage[all]{hypcap}

\newcommand{\C}{^{13} \text{C}}
\newcommand{\fref}[1]{Fig.~\ref{fig:#1}}
\newcommand{\eref}[1]{Eq.~(\ref{eq:#1})}
\usepackage{textcomp}
\usepackage{color}
\usepackage{afterpage}
\usepackage[normalem]{ulem}
\usepackage{pifont}

\usepackage{braket}
\usepackage{natbib}

\begin{document}
\title{Dephasing mechanisms of diamond-based nuclear-spin memories for quantum networks} 
\author{N. Kalb}
\author{P.C. Humphreys}
\author{J.J. Slim}
\author{R. Hanson}
\email{r.hanson@tudelft.nl}
\affiliation{QuTech, Delft University of Technology, P. O. Box 5046, 2600 GA Delft, The Netherlands}
\affiliation{Kavli Institute of Nanoscience, Delft University of Technology, P. O. Box 5046, 2600 GA Delft, The Netherlands}

\begin{abstract}
	We probe dephasing mechanisms within a quantum network node consisting of a single nitrogen-vacancy centre electron spin that is hyperfine coupled to surrounding $\C$ nuclear-spin quantum memories. Previous studies have analysed memory dephasing caused by the stochastic electron-spin reset process, which is a component of optical internode entangling protocols. Here, we find, by using dynamical decoupling techniques and exploiting phase matching conditions in the electron-nuclear dynamics, that control infidelities and quasi-static noise are the major contributors to memory dephasing induced by the entangling sequence. These insights enable us to demonstrate a 19-fold improved memory performance which is still not limited by the electron reinitialization process. We further perform pump-probe studies to investigate the spin-flip channels during the optical electron spin reset. We find that spin-flips occur via decay from the meta-stable singlet states with a branching ratio of 8(1):1:1, in contrast with previous work. These results allow us to formulate straightforward improvements to diamond-based quantum networks and similar architectures.
\end{abstract}

\maketitle

\section{Introduction}

The creation of a general-purpose quantum network will enable distributed quantum computation and long-distance quantum communication~\cite{kimble_quantum_2008}.
A quantum network is composed of individual nodes each hosting a number of qubits that are commonly separated into two groups: communicators and memories.
Communicators have an efficient optical interface that allows for the generation of spin-photon entanglement and ultimately the creation of inter-node entanglement.
Memories on the other hand are robust qubits that allow for intra-node interfacing with the communicators and thus grant access to multi-qubit protocols and the creation of highly-linked many-body quantum states across complex network architectures.\\

To date quantum network primitives have been demonstrated on several experimental platforms by creating point-to-point entangling links between nodes that either were comprised of one communicator each~\cite{moehring_entanglement_2007,ritter_elementary_2012,hofmann_heralded_2012,bernien_heralded_2013,northup_quantum_2014,narla_robust_2016,stockill_phase-tuned_2017} or of one communicator and one additional memory in one of the nodes that would rapidly dephase during internode entanglement generation~\cite{pfaff_unconditional_2014,hucul_modular_2015}.
Very recently, nitrogen-vacancy (NV) centres in diamond have been able to perform network protocols that demand the storage and processing of two entangled states~\cite{kalb_entanglement_2017}, i.e. one communicator was linked with one fully coherent memory (\fref{Fig1}A). 
Nodes combining communicator and memory qubits are readily available in diamond of natural isotopic composition as NV centres are surrounded by a dilute bath of $\C$ nuclear spins ($I=\frac{1}{2}$, $1.1\,\%$ abundance).
Each NV electron spin can selectively address nuclear spins in the near vicinity via dynamical decoupling techniques~\cite{zhao_sensing_2012,kolkowitz_sensing_2012,taminiau_universal_2014} thus making it a natural communicator surrounded by nuclear-spin memories.\\

The NV electron spin and the nuclear spins interact via the always-on magnetic hyperfine interaction. Uncontrolled electron spin flips therefore translate into uncontrolled shifts of the nuclear precession frequency giving rise to nuclear-spin dephasing.
Previous work analysed the impact of stochastic NV reinitialization, a key ingredient for current probabilistic NV-NV entangling sequences, on nuclear-spin decoherence~\cite{blok_towards_2015,reiserer_robust_2016}.
It was implicitly assumed in these works that this constituted the dominant decoherence pathway.
In contrast, here we find that several other mechanisms in fact constitute the dominant sources of decoherence during entanglement generation.
We show that electron spin control errors during entangling attempts in combination with quasi-static noise overshadow dephasing from the NV reinitialization.
These insights enable accurate modelling of the system and as a result uncover direct paths to improved memory robustness through shortened entangling sequences and increased magnetic fields.\\

This work is structured as follows. In Sec.~\ref{sec:setup}, we describe the NV system, the interaction Hamiltonian with surrounding nuclear spins and the sources of nuclear-spin dephasing introduced by repetitive entangling attempts on the electron spin. 
Section~\ref{sec:c1_results} provides evidence for the introduction of additional quasi-static noise during entangling attempts by observing that a nuclear-spin inversion enhances the memory robustness. 
Section~\ref{sec:mw} combines nuclear-spin inversion rotations with time-tailored entangling attempts that render the sequence robust with respect to microwave control errors. 
Using these entangling sequences we investigate two nuclear spin memories and observe an order of magnitude improved memory performance. 
In Sec.~\ref{sec:singlet}, the electron reinitialization process is investigated via nanosecond-resolved pump-probe experiments. 
We quantify into electron spin-flip mechanisms and branching ratios from the meta-stable singlet states to the NV ground state. 
In Sec.~\ref{sec:current_best}, the memory performance is further optimized by exploring the entangling-attempt parameter space. 
We find that the investigated memories are not limited by the stochastic repumping process but rather by a combination of intrinsic decoherence, slowly-fluctuating noise, electron-spin initialization errors and depolarization noise. 
In Sec.~\ref{sec:nv_init}, nuclear-spin dephasing due to electron-spin initialization errors is investigated. 
While noticeable effects of initialization failure are observed, the magnitude of this noise source does not suffice to solely explain the previously observed limitations to nuclear spin memory robustness. 
We conclude with Sec.~\ref{sec:outlook} by inferring favourable parameter regimes from a Monte-Carlo simulation and by suggesting future experimental directions. 

\begin{figure}[tbh!]
	\centering
	\includegraphics[width=8.3cm]{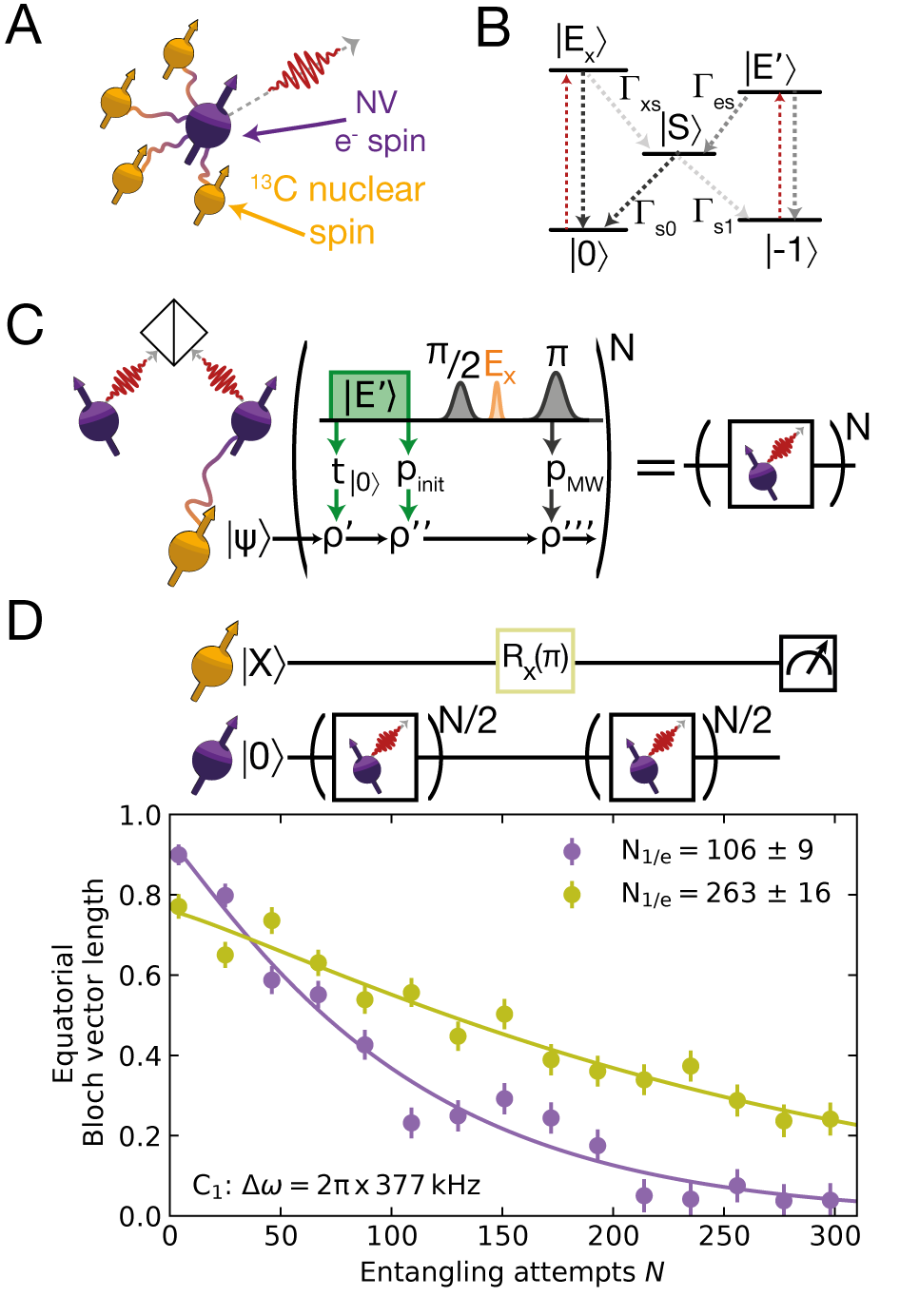}
	\caption{NV centres in diamond as multi-qubit nodes for quantum networks. (\textbf{A}) The NV electron spin (purple) serves as optical interface (red wavepacket) to establish remote entangling links. The surrounding $\C$ nuclear spins (orange) are hyperfine coupled to the electron spin (wiggly lines) and serve as quantum memories. (\textbf{B}) Relevant level structure of the NV centre. The NV allows for spin-selective optical transitions to $\ket{\mathrm{E_x}}$ and $\ket{\mathrm{E'}}$. These states may decay to the meta-stable singlet states summarized as $\ket{S}$. The transition rates $\Gamma_i$ are qualitatively indicated by the opacity of the dashed arrows.  (\textbf{C}) Repeated attempts are made to create remote NV-NV entanglement until success is heralded. Each entangling attempt consists of electron spin manipulations via laser (green/orange) and microwave pulses (grey). Electron spin operations fail with probability $p_i$, thus leaving the electron spin in a mixed state. Nuclear spins with initial state $\ket{\psi}$ will acquire an electron-state-dependent phase which results in a mixed state ($\rho'$, see \eref{h}). Further nuclear decoherence is induced by the optical electron spin reset ($\mathrm{\ket{E'}}$), a stochastic process with randomly distributed projection times $t_{\ket{0}}$. (\textbf{D}) Top: Experimental sequence. Bottom: Memory coherence decay of nuclear spin $\mathrm{C_1}$ ($\Delta \omega=2\pi \times 377\,\mathrm{kHz}$) with (yellow) and without (purple) interleaved $\pi$-rotation to probe quasi-static noise. See legend for fitted decay constants $N_\mathrm{1/e}$. Error bars represent one standard deviation.}
	\label{fig:Fig1}
\end{figure}

\section{Experimental system}
\label{sec:setup}

All experiments were performed on a type IIa chemical-vapor deposition diamond sample that was cut along the $\langle$111$\rangle$ crystal axis and grown by Element Six. 
We milled a solid immersion lens around the positions of single NV centres to enhance photon collection efficiencies and use an additionally grown $\mathrm{Al_2O_3}$ anti-reflection coating~\cite{yeung_anti-reflection_2012}. 
Lithographically-defined gold microstructures allow for the on-chip delivery of amplitude-shaped Hermite microwave pulses for NV electron spin control at a magnetic field of $414\,\mathrm{G}$ aligned with the NV symmetry axis. 
The sample is situated in a home-built cryogenic confocal microscope setup ($T=4$\,K) to allow for resonant single-shot readout of the electron spin state at the $\ket{0} \equiv \ket{m_s=0} \rightarrow \ket{\mathrm{E_x}}$ excitation frequency (\fref{Fig1}B; all shown data throughout this work are corrected for electron read-out infidelities and NV ionization events)~\cite{robledo_high-fidelity_2011,reiserer_robust_2016}.
We use a second laser beam resonant with the optical transitions $\ket{m_s=\pm 1} \equiv \ket{\pm 1} \rightarrow \ket{E'}$ for fast electron spin initialization into the $\ket{0}$ spin state.
Note that the symbol $\ket{E'}$ is used as a shorthand to denote the two optically excited states $\ket{E_{1,2}}$.
Before each experimental run we monitor the NV fluorescence under optical excitation to ensure the required resonance conditions and charge-state occupation~\cite{robledo_high-fidelity_2011}.
We further use time-tailored dynamical decoupling sequences on the NV electron spin to selectively address and control nuclear spins in the vicinity~\cite{taminiau_universal_2014}.\\

Nuclear spins are natural quantum memories due to their long coherence times while the electron spin remains idle~\cite{yang_high-fidelity_2016}.
If, however, states are stored on the nuclei while the electron spin is manipulated then electron-spin control errors can propagate onto the nuclear spin state via the hyperfine interaction. 
The electron-nuclear Hamiltonian in an appropriately rotating frame and secular approximation is
\begin{equation}
\label{eq:h}
H = \omega_0I_\mathrm{z} + (A_\parallel S_\mathrm{z}I_\mathrm{z} + A_\perp S_\mathrm{z}I_\mathrm{x})
\end{equation}
with the nuclear and electronic spin operators $I_j$ and $S_j$, the bare Larmor frequency arising from the external magnetic field $\omega_0 = 2\pi \gamma |\vec{B}| = 443275\,\mathrm{Hz}$ and the parallel (perpendicular) hyperfine coupling strength $A_\parallel$ ($A_\perp$).
The Hamiltonian of \eref{h} gives rise to electron-spin-dependent nuclear precession frequencies $\omega_0$ (electron spin in $\ket{0}$) and $\omega_{\pm1} = \sqrt{(\omega_0 \pm A_\parallel)^2+A_\perp^2}$ (electron in $\ket{m_s = \pm 1}$).
It is therefore evident that unaccounted electron spin flips will decohere nuclear-spin memories due to a shift in nuclear precession frequency $\Delta \omega = |\omega_0-\omega_{\pm1}|$.\\

The generation of long-distance entangled states is a central source of electron spin state uncertainty and thus nuclear spin decoherence (\fref{Fig1}C).
Independent of the entangling scheme~\cite{cabrillo_creation_1999,barrett_efficient_2005,campbell_measurement-based_2008}, each entangling attempt involves error-prone operations such as microwave spin rotations, optical excitation ($\mathrm{E_x}$) and optical spin reinitialization into $\ket{0}$ ($\mathrm{E'}$).
The timing of these operations is typically optimized to preserve the coherence of both the nuclear and electron spin~\cite{reiserer_robust_2016}.
Any residual imperfections in these operations will however give rise to dephasing of nuclear spin superposition states.
Moreover the electron spin reinitialization relies on optical pumping, an inherently stochastically-timed process that poses a limit on the number of entangling attempts a nuclear spin of a certain coupling strength $\Delta \omega$ can preserve a quantum state for.
In the experiments described here we use entangling sequences that contain all necessary electron-spin operations apart from the generation of spin-photon entanglement ($\mathrm{E_x}$). 
We expect the impact of undesired electron spin flips ($p \approx 0.005$) after an optical excitation to $\mathrm{E_x}$ to be negligible when compared to other sources of error because the timing of these optical excitation pulses can be chosen such that the spurious nuclear spin phases upon flipping are small~\cite{reiserer_robust_2016}.\\

\section{Performance of a strongly-coupled nuclear spin memory}
\label{sec:c1_results}
We first examine the coherence of nuclear spin $\mathrm{C_1}$, a particularly strongly coupled nuclear spin in close proximity to our NV centre (Table~\ref{tab:13Cs} lists the key numbers for seven addressable carbon nuclear spins). 
The spin is initialized into the superposition state $\ket{X} \equiv (\ket{\uparrow}+\ket{\downarrow})/\sqrt{2}$ and after a number of entangling attempts (duration of one attempt: $7\,\mathrm{\mu s}$), the nuclear spin coherence is measured by evaluating the remaining length of the Bloch vector in the equatorial plane of the Bloch sphere $\sqrt{\langle \sigma_\mathrm{x}\rangle^2+\langle \sigma_\mathrm{y}\rangle^2}$ (\fref{Fig1}D), with the Pauli spin operators $\sigma_\mathrm{i}$. 
We find an exponential decay of the nuclear coherence with a $1/e$ decay constant of $106(9)$ attempts for a consecutive stream of entangling attempts (purple data in \fref{Fig1}D).
By further employing a Hahn-echo $\pi$ rotation ($\mathrm{R_x(\pi)}$) on the nuclear spin after half the attempts to cancel quasi-static noise, we obtain an improved decay constant of $263(16)$ attempts (yellow data). 
Here $\mathrm{R_i(\theta)}$ corresponds to a rotation around axis $i$ with angle $\theta$.
These coherence decays are remarkable since the coupling strength $\Delta \omega$ for this spin is an order of magnitude larger than the nuclear spin memories used for the recent demonstration of entanglement distillation, yet a comparable decoherence rate is observed~\cite{kalb_entanglement_2017}.
These results therefore provide a key first indication that the model of Refs.~\cite{blok_towards_2015,reiserer_robust_2016} does not fully capture the decoherence dynamics of the NV-nuclear system.\\ 

\begin{table}
	\centering
	\begin{tabular}{c|c|c|c|c|c|c|c}
		
		&  C$_1$ & C$_2$	 & C$_3$ & C$_4$	 & C$_5$	 & C$_6$	 & 	C$_7$	 \\ 
		\hline
		\hline
		$\frac{\Delta \omega}{2\pi}\,(\mathrm{kHz})$	 & 376.5  &  62.4 & 77.0  & 32.4 & 26.6 & 20.9 & 12.2  \\
		\hline
		$T_2^*\,(\mathrm{ms})$ & 9.9(2) & 9.9(1) & 9.5(2) & 11.2(3) & 17.3(6)& 4.5(1) & 7.0(1) \\
	\end{tabular}
\caption{Coupling strength $\Delta\omega$ and free-induction decay $T_2^*$ for seven addressable nuclear spins in the vicinity of the NV. Note that the NV used in this work does not correspond to the NV used in Ref.~\cite{reiserer_robust_2016}. \label{tab:13Cs}}
\end{table}
The increased $1/e$ decay constant for an interleaved nuclear $\mathrm{R_x(\pi)}$ rotation points towards quasi-static noise introduced by the repeated performance of entangling attempts as a large component of the nuclear spin decoherence. 
Such quasi-static noise may originate from slow intensity fluctuations of the repumping laser at the position of the NV which may, for example, be induced by mechanical vibrations of the optical set-up. 
Intuitively, the distribution of electron-spin-reset times will fluctuate in accordance with the laser intensity at the NV-position. 
Fluctuating electron-spin-reset times directly translate into a fluctuation of the average phase per entangling attempt imprinted onto the nuclear spins~\cite{reiserer_robust_2016}.
In the future, the exact origin of the quasi-static noise could be probed by measuring the pointing stability of the impinging beam with respect to the NV and employing active laser-intensity stabilization methods. \\
The coupling strength $\Delta\omega$ and nuclear decoherence due to NV reinitialization are related via the model of Blok et al.~\cite{blok_towards_2015}.
In this model a decay constant $\tau$ is invoked to model the decoherence of nuclear spins with a given coupling strength $\Delta \omega$.
This model will further apply to any noise source that induces stochastic noise in the timing of the NV spin population.
Note that $\tau$ becomes the mean of the exponentially-distributed electron-spin repumping time under the assumption that this repumping process is the dominant noise source. 
The expected equatorial Bloch vector length is a function of $\tau$, $\Delta\omega$, the number of entangling attempts $N$ and the probability $p_\mathrm{\ket{1}}$ that the electron spin is in $\ket{\pm 1}$ at the end of the entangling attempt
\begin{equation}\label{eq:blok}
\sqrt{\langle \sigma_\mathrm{x}\rangle^2+\langle \sigma_\mathrm{y}\rangle^2} = (1-p_\mathrm{\ket{1}}+p_\mathrm{\ket{1}}e^{-\Delta \omega^2 \tau^2/2})^N.
\end{equation}
Using \eref{blok} we obtain $\tau \approx 52\,\mathrm{ns}$ for the best dataset (\fref{Fig1}D, yellow), which is a factor of 8 faster and therefore at odds with the observed data in Ref.~\cite{reiserer_robust_2016}.
We hypothesize that this discrepancy arises as $\tau$ originates not only from the reinitialization process, but is also impacted by other NV control infidelities. 
This motivates the further investigations in the following sections.\\

\section{Microwave control errors}
\label{sec:mw}
We next consider the influence of microwave control errors and their impact on the nuclear spin coherence.
All entangling sequences contain a microwave-induced $\pi$-rotation to preserve the electron coherence upon successfully creating entanglement and to render the acquired phase of the nuclear spin electron-state-independent.
This is because the time spent in both electron states equalizes for a perfect $\pi$-rotation and the nuclear spin picks up a phase according to the average frequency $\bar{\omega}=(\omega_0+\omega_{-1})/2$ and the inter-pulse delay $t$.
Microwave $\pi$-rotations however fail with a probability $p_\mathrm{MW}$ such that the rotation can be described as a mixture of two processes $p_\mathrm{MW}\times \mathrm{R_x(0)} + (1-p_\mathrm{MW})\times \mathrm{R_x(\pi)}$.
The nuclear spin therefore acquires a phase $\varphi$ depending on the success or failure of the microwave $\pi$ pulse and the projected electron state after the initial $\pi/2$ microwave rotation (\fref{Fig2}). The probabilities : phases for the different possible outcomes are
\begin{equation} \label{eq:phase_mw}
\begin{split}
	(\ket{0};\ket{0})\quad 0.5\,p_\mathrm{MW}:\,& \varphi_0 = 2\omega_0 t\\
	(\ket{-1};\ket{-1})\quad 0.5\,p_\mathrm{MW}:\,& \varphi_1 = 2\omega_{-1} t\\
	(\ket{0};\ket{-1})\quad 1-p_\mathrm{MW}:\,& \bar{\varphi} = 2\bar{\omega} t.
\end{split}
\end{equation}
We additionally list the electron spin state in brackets ($\mathrm{\ket{before};\ket{after}}$) before and after the pulse has been applied.
The inter-pulse delay $t$ can be chosen such that the acquired phase equalizes in all cases $\bar{\varphi}-\varphi_0\,(\mathrm{mod}\,\,2\pi)=\bar{\varphi}-\varphi_1\,(\mathrm{mod}\,\,2\pi)$.
This condition is fulfilled for $t = 2\pi/\Delta \omega$.\\

\begin{figure}[tbh!]
	\centering
	\includegraphics[width=\columnwidth]{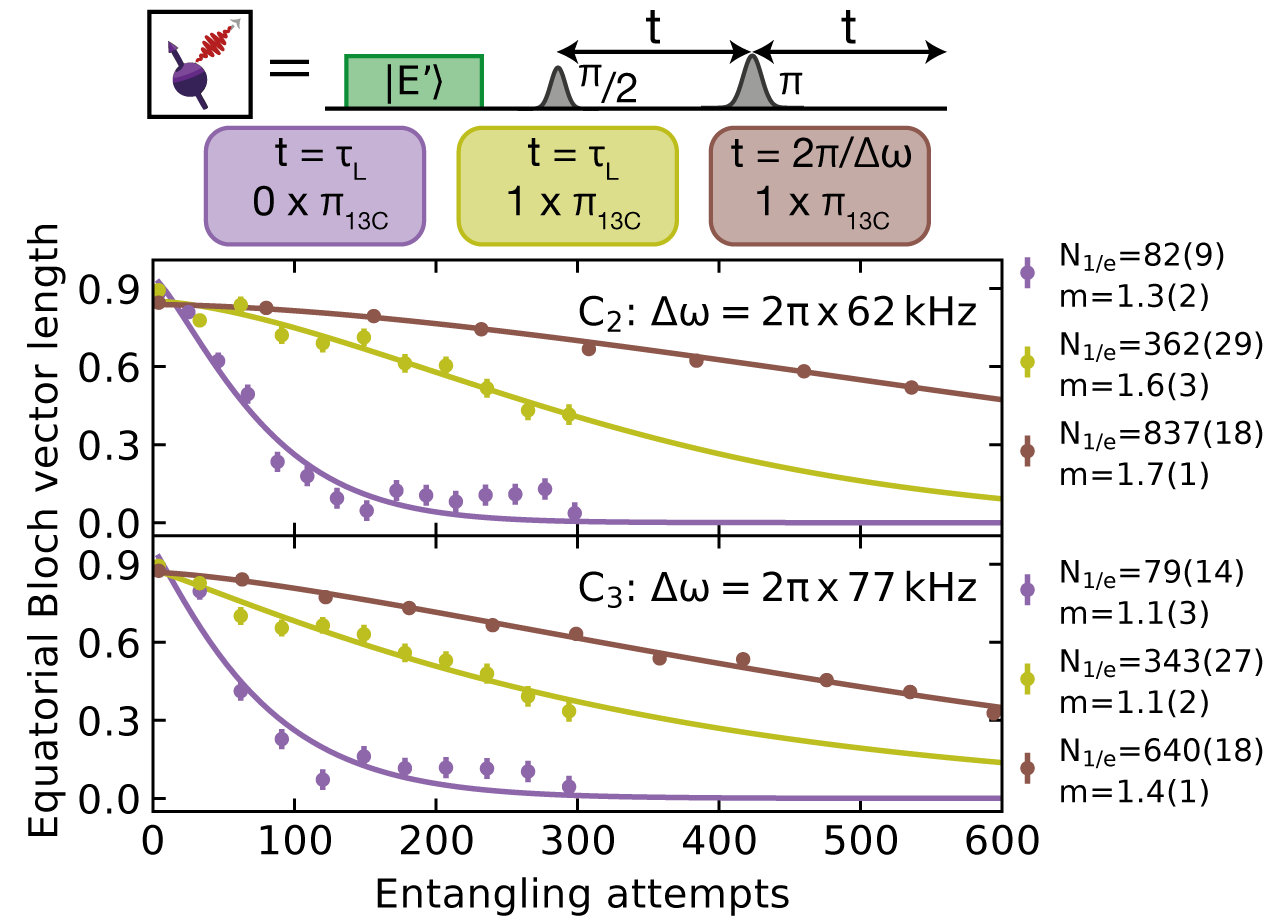}
	\caption{Impact of microwave pulse errors on memory performance. We measure the decay of a superposition state on two different nuclear spins $\mathrm{C_2}$, $\mathrm{C_3}$ for three different scenarios (see legend and main text). We find maximal decay constants of 837(18) and 640(18) respectively which corresponds to $\tau = 177\,\mathrm{ns}$/$163\,\mathrm{ns}$ according to Blok et al.~\cite{blok_towards_2015}. Solid lines are exponentially decaying fits to the data sets with $A\times \mathrm{Exp}[-(N/N_{1/e})^m]$. Error bars are one s.d.}
	\label{fig:Fig2}
\end{figure}

We use these phase-matched entangling attempts to unveil the impact of microwave pulse errors.
We choose two nuclear spins with intermediate coupling strengths ($\mathrm{C_2}$: $\Delta\omega =2\pi\times62\,\mathrm{kHz}$ and $\mathrm{C_3}$: $\Delta\omega =2\pi\times77\,\mathrm{kHz}$ respectively) and measure their decoherence rate in three different scenarios (\fref{Fig2}).
First, the standard setting of Ref.~\cite{reiserer_robust_2016} with an inter-pulse delay corresponding to the inverse of the nuclear spin Larmor frequency $t=\tau_\mathrm{L} = 2\pi/\omega_0$ (purple).
Second, with a single interleaved nuclear $\pi$-rotation to overcome quasi-static sources of dephasing and $t=\tau_L$ (yellow).
And finally with an optimized inter-pulse delay $t=2\pi/\Delta\omega$ (\eref{phase_mw}, brown).
All data are fit with an exponentially decaying function $A\times \mathrm{Exp}[-(N/N_{1/e})^m]$ with free parameters $A$, $N_{1/e}$ and $m$.\\

We find an increase in robustness by an order of magnitude when comparing the optimized sequence (brown) to the standard sequence (purple).
In addition, the optimized sequence outperforms the standard sequence with inversion rotation (yellow) therefore demonstrating the role of microwave pulse errors during the entangling sequence.
The coherence decay of $\mathrm{C_2}$ and $\mathrm{C_3}$ deviates from the  exponential decay observed for $\mathrm{C_1}$ and predicted by Ref.~\cite{blok_towards_2015}. 
We still use \eref{blok} to associate the observed $1/e$-decay constants with $\tau$ therefore establishing a basis for performance comparisons between nuclear spins.
We find $\tau = 177\,\mathrm{ns}$ ($163\,\mathrm{ns}$) for $\mathrm{C_2}$ ($\mathrm{C_3}$) which is still larger than the earlier obtained value of $52\,\mathrm{ns}$ for $\mathrm{C_1}$. 
We note that $\mathrm{C_1}$ did not show a changed decay behaviour for the optimized entangling sequence ($N_{1/e} = 265(28)$ and $m = 1.0(2)$; data not shown) leading us to conclude that the dominant noise source of $\mathrm{C_1}$ is indeed the stochastic electron spin reinitialization.
These elevated decay constants for $\mathrm{C_2}$ and $\mathrm{C_3}$ require a deeper understanding of the NV electron spin repumping process and the exploration of the entangling attempt parameter space to yield further improvements in memory robustness.\\

\section{Spin-flip mechanisms of the NV electron spin}
\label{sec:singlet}

We next investigate spin-flip mechanisms of the NV electron spin during the optical pumping process. 
Spin-flips are expected to either be direct, i.e. they occur via spin-mixing in the excited state, or indirect via NV specific intersystem crossing (ISC) to the orbital singlet states (summarized here as $\ket{\mathrm{S}}$; \fref{Fig1}B). 
The electron spin then decays from $\ket{\mathrm{S}}$ back to the NV spin ground-state triplet of $m_s = 0$ and $ m_s=\pm1$ with the branching ratio $\Gamma_\mathrm{s0}/\Gamma_\mathrm{s1}:1:1$~\cite{doherty_nitrogen-vacancy_2013}. \\

We prepare the NV centre in $\ket{-1}$ and use a calibrated optical $\pi$ pulse to excite the NV to $\ket{\mathrm{E'}}$ with a Gaussian intensity envelope that has a full width at half maximum of $2.6\,\mathrm{ns}$.
We next apply a 40 ns long optical pulse on the $E_\mathrm{y}$ transition to detect fluorescence of the $\ket{0}$ ground-state and therefore spin flips during the optical cycle.
By varying the delay between both laser pulses we are able to monitor spin flips to $\ket{0}$ in a time-resolved fashion.
The data set is generalized by using DC Stark tuning~\cite{tamarat_stark_2006} to induce a range of frequency shifts of $\ket{\mathrm{E_x}}$ ($1-4.5\,\mathrm{GHz}$) in the NV excited state.
All measurements are normalized to interleaved experimental runs where no optical $\pi$ pulse is applied and the NV is either prepared in $\ket{0}$ or $\ket{-1}$.\\

We observe an exponential increase of the probability to be in $\ket{0}$ with a strain-averaged time scale of $368(12)\,\mathrm{ns}$ (\fref{Fig3a}) which is consistent with the literature value for the singlet lifetime of single NV centres $371\,\mathrm{ns}$~\cite{robledo_spin_2011} and in reasonable agreement with ensemble measurements~\cite{acosta_optical_2010}. 
Only a negligible fraction of the spin population resides in $\ket{0}$ for short pump-probe delays, therefore ruling out direct decay from $\ket{E'}$ to $\ket{0}$ as dominant electron spin-flip mechanism.
Since the probe window has a finite length (40 ns) the observed direct spin-flip probability ($1(1)\,\%$) represents an upper bound.
We are therefore able to identify the decay from $\ket{S}$ to the ground state triplet as the major spin-flip mechanism. 
The measured singlet lifetime allows us to estimate the expected decoherence if the singlet states were to couple significantly to the nuclear spins. 
In this case, we would expect much faster spin decay than observed~\cite{blok_towards_2015}, allowing us to rule out such an effect as a significant decoherence mechanism in our experiments.\\

The data in \fref{Fig3a} allow for the determination of the spin-flip probability per optical excitation to $\ket{\mathrm{E'}}$ and --- together with the measured ISC rate $\Gamma_\mathrm{es}$ --- the branching ratio from the singlet states $\ket{\mathrm{S}}$ to the NV ground state.
Following the methods of Ref.~\cite{goldman_phonon-induced_2015-1} we experimentally determine the ISC rate $\Gamma_\mathrm{es}$ for the $\ket{\mathrm{E'}}$ states. We quote strain-averaged values as we assume no significant strain dependence (all fitted values are given in Table~\ref{tab:singlet}).
By measuring the radiative lifetime of $\ket{\mathrm{E_x}}$ and $\ket{\mathrm{E'}}$ via resonant optical excitation and time-resolved fluorescence monitoring in the phonon sideband we obtain strain-averaged lifetimes $t_\mathrm{Ex} = 12.3(1)\,\mathrm{ns}$ and $t_\mathrm{E'} = 7.4(1)\,\mathrm{ns}$ (\fref{Fig3a} inset).
From the measured lifetimes and the assumption that the transition rate from $\mathrm{\ket{E_x}}$ to $\mathrm{\ket{S}}$ is $\Gamma_\mathrm{xs} \approx 0$~\cite{goldman_phonon-induced_2015-1}, we extract a strain-averaged ISC rate for $\mathrm{E'}$ of $\Gamma_\mathrm{es} = 2\pi \cdot 8.2(2)\,\mathrm{MHz}$ in good agreement with earlier results~\cite{goldman_phonon-induced_2015-1}.\\

From the measurements described in this section ($\Gamma_\mathrm{es},t_\mathrm{Ex},t_\mathrm{E'}$) we obtain a strain-averaged probability of $p_s = 0.41(1)$ to transfer to $\ket{S}$ per excitation cycle on $\ket{\mathrm{E'}}$.
The probability for double excitation to $\ket{\mathrm{E'}}$ may obscure the estimate of the singlet branching ratio~\cite{humphreys_deterministic_2017}.
We use a quantum jump simulation to estimate this probability with $t_\mathrm{E'}$, $t_\mathrm{Ex}$ and the intensity profile of the excitation pulse as input parameters~\cite{humphreys_deterministic_2017}.
Our simulation results in a double excitation probability of $\sim 5\,\%$, which we take into account when computing $p_s$.\\

From $p_s$, the measured probability to be in $\ket{0}$ after one excitation and the assumption that the decay rate from $\ket{S}$ to $\ket{\pm1}$ is symmetric~\cite{reiserer_robust_2016} we are able to extract the branching ratios from $\ket{S}$ to $\ket{0}$:$\ket{+1}$:$\ket{-1}$. 
We find a strain-averaged branching ratio of $8(1)$:$1$:$1$. This value is significantly different from the literature value (2:1:1) for NV centres at ambient temperatures~\cite{doherty_nitrogen-vacancy_2013}. 
The relatively large uncertainty originates from the spread in $F(\ket{0})$ for long delay times and may be explained by imperfect optical $\pi$ excitations. 
We emphasize that the lowest extracted branching ratio is $5(1):1:1$ while the largest obtained branching ratio is $13(3):1:1$ therefore validating a data set that is systematically above the literature value (see Table~\ref{tab:singlet}). 
This result is in accordance with Ref.~\cite{felton_hyperfine_2009} which found a higher spin polarization at cryogenic temperatures upon off-resonant excitation. 
Note that the observed branching ratio also suffices to explain the electron spin reinitialization data of Ref.~\cite{reiserer_robust_2016} without invoking direct spin-flip channels. 
In the future, the experimental tools developed in this section can be used to study the temperature dependence of the singlet branching ratio on single NV centres.\\

\begin{table}
	\centering
	\begin{tabular}{c|c|c|c}
		
		$\Delta_\perp\,(\mathrm{GHz})$	& lifetime (ns) & $p_s$ & $\ket{0}$:$\ket{+1}$:$\ket{-1}$\\
		\hline
		\hline
		0.9	 & 379(17) &  0.41(1) & 11(2):1:1\\
		1.3	 & 340(18) & 0.39(1) &13(3):1:1\\
		1.7	 & 403(26) & 0.39(2)&5(1):1:1 \\
		2.7	 & 343(32) & 0.43(2)&5(1):1:1\\
		4.6	 & 372(37) & 0.42(2)&6(1):1:1\\
	\end{tabular}
	\caption{Results of the singlet pump-probe experiments. Ordered according to the electric-field-induced frequency shift $\Delta_\perp$ of $\ket{\mathrm{E_x}}$. We give the inferred cumulative singlet lifetime, the probability of transfering to the singlet $p_s$ after one excitation pulse and the branching ratio from the singlet states into the ground-state spin triplet. \label{tab:singlet}}
\end{table}

\begin{figure}[tb!]
	\centering
	\includegraphics[width=7.5cm]{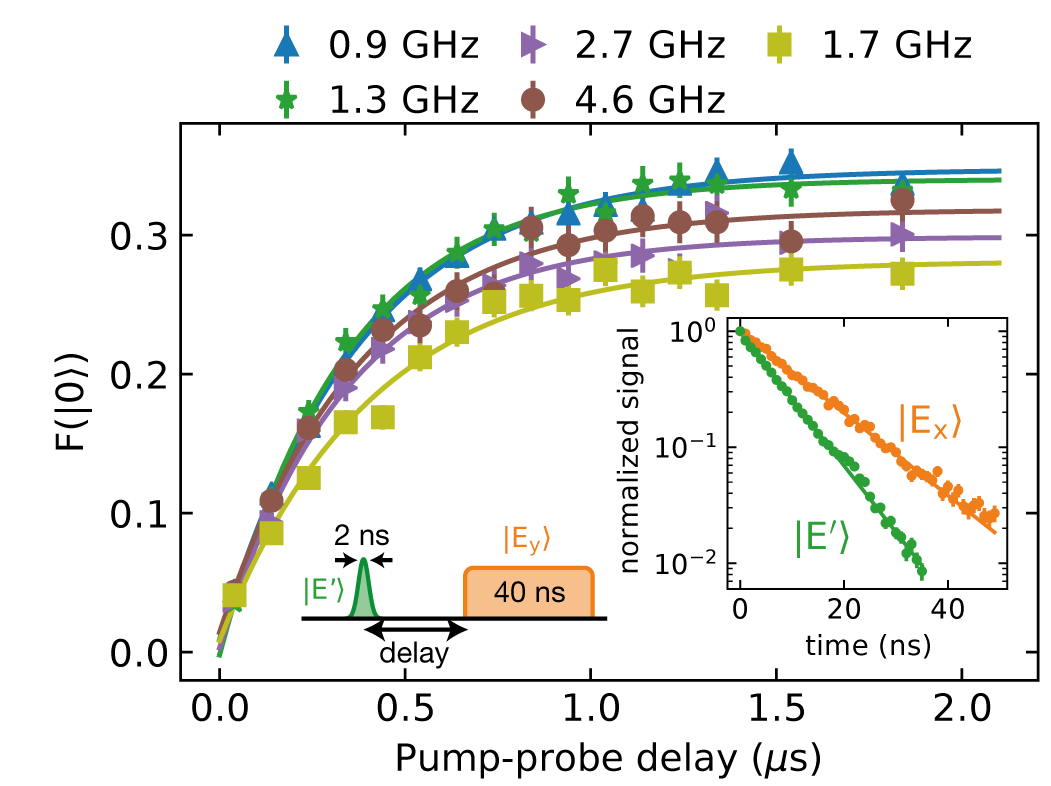}
	\caption{Results of a pump-probe experiment on the $\ket{-1} \rightarrow \ket{\mathrm{E'}}$ transition. We measure the radiative lifetimes of both states (inset) and infer $\Gamma_\mathrm{es}/2\pi = 8.2(2)\,\mathrm{MHz}$. Besides, we infer from this a strain-averaged singlet branching ratio of $8(1):1:1$. See legend for the strain-induced frequency shifts at which each data set was taken.
	}
	\label{fig:Fig3a}
\end{figure}

\section{Current limits to memory robustness}
\label{sec:current_best}
We investigate factors that currently limit the nuclear spin coherence for spins $\mathrm{C_2}$ and $\mathrm{C_3}$. 
To this end we initialize the nuclear spin in a balanced superposition and measure the remaining coherence $\sqrt{\langle \sigma_\mathrm{x}\rangle^2+\langle \sigma_\mathrm{y}\rangle^2}$ after a number of entangling attempts while changing several key attributes of the repeated electron-spin entangling sequence.
First, we remove the intermediary electron $\mathrm{R_x(\pi)}$ rotation and set the time after the first microwave pulse ($\mathrm{R_x(\alpha)}$) to $t \sim 2\pi/\Delta\omega$ (\eref{phase_mw} and \fref{Fig3b}) which removes the dependency of the acquired nuclear spin phase on the electronic spin state. 
This allows us to effectively half the entangling attempt duration while still applying electron spin reinitialization events repetitively.
Second, we sweep the intensity of the optical pumping beam to obtain the influence of the electronic reinitialization speed on the nuclear spin coherence for each data set.
In addition, the influence of quasi-static noise is further probed by either interleaving a single (purple data) or two (yellow data) nuclear spin inversions at an appropriate timing.
Note that all previous experiments were conducted with a reinitialization intensity of $6\,\mathrm{\mu W}$ and a repumping duration of $2\,\mathrm{\mu s}$. \\

\begin{figure}[tb!]
	\centering
	\includegraphics[width=7.5cm]{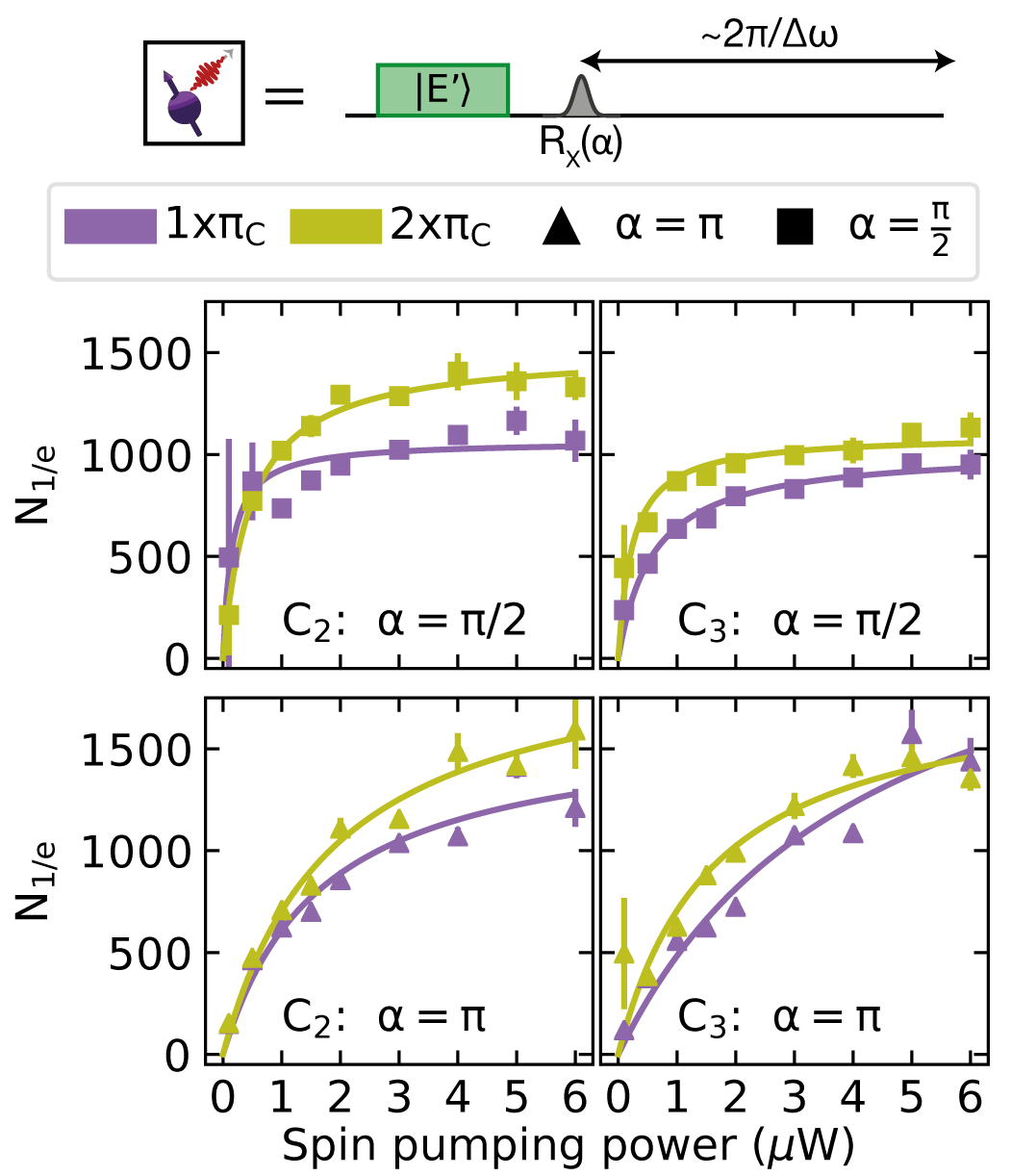}
	\caption{Nuclear spin decay constants as a function of optical spin-pumping power and different sequence configurations. Purple (yellow) data uses one (two) nuclear spin $\pi$ rotations to mitigate quasi-static noise. The microwave pulse during each entangling sequence rotates the electron by an angle $\alpha$ which we set to $\alpha = \pi/2$ (squared, top panels) or $\alpha=\pi$ (triangles, bottom panels). Solid lines are fits to the function $N_\mathrm{sat}\cdot P/(P+P_\mathrm{sat})$ with the optical power $P$ and the free parameters $N_\mathrm{sat}$ and $P_\mathrm{sat}$.
	}
	\label{fig:Fig3b}
\end{figure}

Figure~\ref{fig:Fig3b} shows the inferred decay constants $N_{1/e}$ when fitting the data with the function $A\times \mathrm{Exp}[-(N/N_{1/e})^m]$ with free parameters $A$, $N_{1/e}$, $m$.
We use entangling attempts with $\alpha=\pi/2$ to set $p_{\ket{1}}= 0.5$ (top panels, squares) and find that using two nuclear $\pi$-rotations (yellow data) outperforms the use of a single $\pi$-rotation (purple data). 
This indicates that the previously identified quasi-static noise is not fully mitigated by a single $\pi$ rotation on the probed timescales of $\sim 20\,\mathrm{ms}$.\\

We fit saturation curves $N_\mathrm{sat}\cdot P/(P+P_\mathrm{sat})$ with the optical pumping power $P$ and the free parameters $N_\mathrm{sat}$ and $P_\mathrm{sat}$ to the measured decay constants (solid lines).
The average fitted saturation power of $P_\mathrm{sat} = 366(68)\,\mathrm{nW}$ is consistent with measurements that were directly carried out on the electron spin under similar experimental conditions~\cite{reiserer_robust_2016}. 
The best achievable $N_{1/e}$ decay constants for $\alpha = \pi/2$ are $N_\mathrm{sat,C_2} = 1511(38)$ and $N_\mathrm{sat,C_3} = 1097(40)$. This yields a 19-fold increase in memory robustness for $\mathrm{C_2}$ when compared to the standard performance in \fref{Fig2} (purple data).\\

We next choose $\alpha = \pi$ (preparing the NV in $\ket{-1}$) such that optical electron spin initialization occurs after each entangling attempt, thus amplifying the phase noise due to the reinitialization process.
Repeating the measurements as described above for both nuclear spins (\fref{Fig4}, lower panels) results in an average saturation power of $P_\mathrm{sat} = 2.4(8)\,\mathrm{\mu W}$ and maximal $N_{1/e}$ decay constants of $N_\mathrm{sat,C_2} = 2045(136)$ and $N_\mathrm{sat,C_3} = 2207(278)$.\\

The measured ratios of $N_\mathrm{sat}$ for $\alpha=\pi/2$ and $\alpha=\pi$ are $1.35$ ($\mathrm{C_2}$) and $2.01$ ($\mathrm{C_3}$). 
These ratios are inconsistent with \eref{blok} which predicts a ratio of $N_\mathrm{sat,\pi}/N_\mathrm{sat,\pi/2}=0.5$.
The increased performance of the nuclear spins when setting $\alpha = \pi$ therefore provides further evidence that the memory decay is not dominated by the stochastic reinitialization process.
The discrepancy in saturation powers between the two data-sets will spur further investigations.\\ 

Entangling sequences using electron $\pi/2$ rotations may cause additional noise due to off-axis rotations of the nuclear spin~\cite{reiserer_robust_2016} and the lack of a frozen core if the electron spin is in $\ket{0}$ thus allowing for resonant nuclear-nuclear flip-flop events that translate to lower nuclear coherence times~\cite{zhong_optically_2015}.
Depolarizing noise is investigated by initializing both nuclear spins in the eigenstate $\ket{\uparrow}$ and setting $\alpha=\pi/2$. We obtain decay constants for these eigenstates of $\gtrsim 3500$.
Environmental dephasing due to the fluctuating spin bath and off-resonant pumping from $\ket{0}$ to $\ket{\pm1}$ is investigated by measuring the nuclear coherence decay $T_2$ without microwave pulses in the entangling attempts. We obtain $T_2 \gtrsim 3000$ entangling attempts for both spins at the highest spin pumping powers.
We additionally point out that the fitted average exponent for the two nuclear spins differs [$m_\mathrm{C_2} = 1.89(0.14)$ and  $m_\mathrm{C_3} = 1.49(0.03)$] which implies differing dominant decay mechanisms for each nuclear spin. 
Note that the entangling sequences of this section are suitable for single-photon entangling protocols~\cite{kalb_entanglement_2017,humphreys_deterministic_2017} because the entangling sequence duration is well within $T_\mathrm{2,Hahn}$ of the electron spin; the required electron $\pi$-rotation can be applied once the entanglement generation is successfully heralded.\\

The observed decay constants in this section are in-line with the best-performing decoherence-protected subspaces formed from two nuclear spin memories therefore indicating that the robustness of these low-coupling ($\Delta\omega \sim \,\mathrm{kHz}$) subspaces may be further improved by more than an order of magnitude~\cite{reiserer_robust_2016}. 
These results further establish that the majority of addressable nuclear spins are suitable quantum memories for quantum networks.\\

\section{Electron spin initialization errors}
\label{sec:nv_init}
The nuclear spin coherence may be restricted by unsuccessful electron reinitialization attempts which we explore in this section.
As shown in Sec.~\ref{sec:singlet}, reinitialization of the electron spin occurs by optical pumping of the states $\ket{\pm 1}$ to the intermediate singlet states $\ket{S}$ from where the electron spin either decays to the final state $\ket{0}$ or with equal probability to the other spin states $\ket{\pm 1}$, thus repeating the pumping cycle.
Optical pumping is a stochastic process that, when applied for a finite duration of time, is accompanied by a failure probability $p_\mathrm{init}$ with which the NV electron spin is either left in $\ket{+1}$ or $\ket{-1}$. \\

We probe electron spin initialization failure by running 700 entangling attempts with a $\mathrm{R_x(\pi/2)}$ rotation on the electron spin that is timed such that it fulfils the earlier discovered phase matching condition of $t=2\pi/\Delta \omega$ (\fref{Fig4}A).
We further vary a waiting time $T$ between the end of the repumping laser pulse and the microwave rotation on the electron spin. 
The absolute change in precession frequency of a nuclear spin $|\Delta \omega|$ is almost identical for both electron states $\ket{\pm 1}$, as can be seen from the Hamiltonian in \eref{h} and the fact that we operate in the regime $(\omega_0 \pm A_\parallel)^2 \gg A_\perp^2$.
This means that there is a phase cancellation condition for $T$ at which electron initialization failure does not invoke a phase shift on the nuclear spin.
The repumping duration $t_r$ was chosen to be $2\,\mathrm{\mu s}$ and the optical pumping power was $4\,\mathrm{\mu W}$. 
The nuclear spins are initialized in $\ket{X}$ and inverted midway through the sequence. Finally the quantity $\sqrt{\langle \sigma_\mathrm{x}\rangle^2+\langle \sigma_\mathrm{y}\rangle^2}$ is measured to evaluate the remaining nuclear spin coherence.\\

\begin{figure}[tb!]
\centering
\includegraphics[width=7cm]{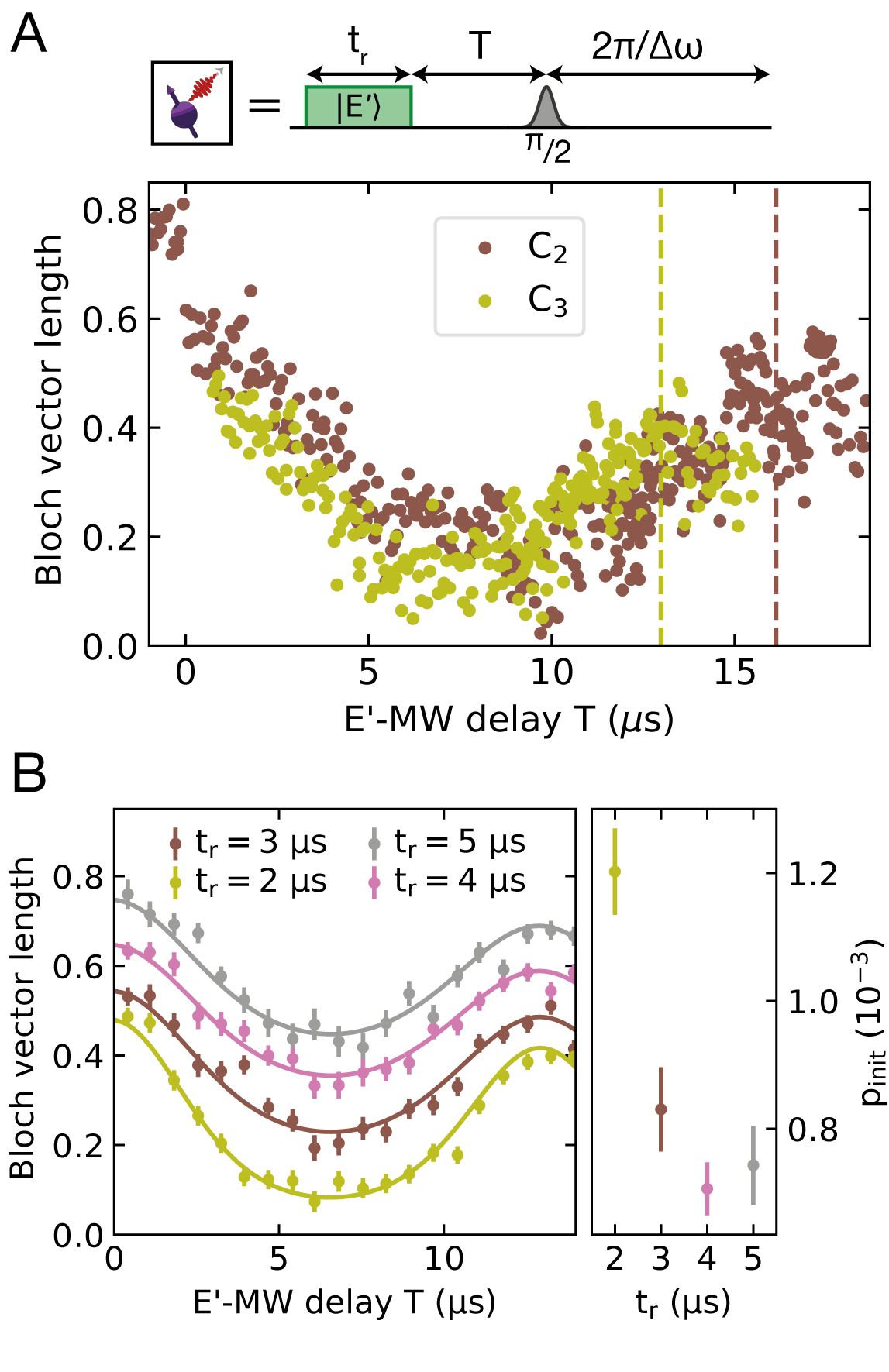}
\caption{\textbf{Memory sensitivity with respect to NV initialization errors.} (\textbf{A}) Top: Entangling sequence performed on the NV electron spin while the nuclear spin is idling in a superposition state. Bottom: Measured equatorial Bloch vector length as a function of the delay $T$ between repumping pulse and microwave $R_x(\frac{\pi}{2})$ rotation. We use $t_r=2\,\mathrm{\mu s}$. Dashed lines are the expected phase matching conditions $T=2\pi/\Delta\omega$ for the respective spin (see legend). (\textbf{B}) Left: Coherence of $\mathrm{C_3}$ for various spin pumping durations $t_r$ (see legend). Solid lines are fits with the probability of initialization failure $p_\mathrm{init}$ as free parameter. The data have been offset for better visibility. Right: Extracted initialization infidelity as a function of repumping duration. Rate calculations suggest that a minimum of $3.3\,\mathrm{\mu s}$ repumping is required to achieve an infidelity of $10^{-4}$ under the assumption of fully saturating the NV and negligible off-resonant excitation. }
\label{fig:Fig4}
\end{figure}

Figure~\ref{fig:Fig4}A shows the measured Bloch vector lengths as a function of $T$.
We observe a decrease in Bloch vector length and a revival of the nuclear spin coherence for both examined nuclear spins (yellow and brown data).
Both nuclear spins experience a revival in coherence at the expected phase matching condition $T = 2\pi /\Delta\omega$ (dashed lines).
The data at negative delays were used to calibrate the delay between the end of the repumping pulse and the start of the microwave rotation on the electron spin (duration of $50\,\mathrm{ns}$).\\

This technique allows us to estimate the initialization failure probability $p_\mathrm{init}$. We measure the nuclear spin coherence of $\mathrm{C_3}$ for four different repumping durations $t_r$ and a repumping power of $4\,\mathrm{{\mu W}}$ (\fref{Fig4}B). The nuclear-spin expectation values $\sigma_\mathrm{x,y}$ after $N$ entangling attempts are described by assuming binomially distributed initialization failures that occur with probability $p_\mathrm{init}$. Each failure is assumed to have equal probability for the electron spin to end up in $\ket{\pm1}$. We use the shorthand notation $b_{mn}(p) = \binom{m}{n} (1-p)^{m-n}p^n$ which results in
\begin{equation}
\begin{split}
\langle \sigma_\mathrm{x}\rangle = \,& A\sum^N_{i=0}b_{Ni}(p_\mathrm{init})\sum_{j=0}^{i}b_{ij}(\frac{1}{2}) \cdot \\
 \,& \cdot \cos\left[ (N-i)\phi_0+j\phi_{-1}+(i-j)\phi_{+1}\right]\\
\end{split}
\end{equation}
where the phases $\phi_{+1,-1,0}$ solely depend on the nuclear spin frequencies for the respective electron spin state and the timing of the used entangling attempts. Analytically evaluating this expression allows us to derive a fit function with two free parameters: $p_\mathrm{init}$ and an amplitude $A$ that encompasses other sources of infidelity (solid lines). The best fitted initialization failure probability is $p_\mathrm{init} = (7.1\pm 0.4 )\cdot10^{-4}$ for $t_r = 4\,\mathrm{\mu s}$ (\fref{Fig4}B, right panel). We attribute the saturation of $p_\mathrm{init}$ for longer repumping durations to off-resonant optical excitation of the NV electron spin.\\

The inferred initialization failure probabilities are used to find an upper-bound to the coherence decay constants of $\mathrm{C_2}$ and $\mathrm{C_3}$ (see \ref{sec:current_best}). 
Using a Monte-Carlo simulation of the nuclear-electron dynamics and the assumptions $p_\mathrm{MW} = 0$, negligible quasi-static noise and an average spin reinitialization time of $52\,\mathrm{ns}$ we find: $N_{1/e,\mathrm{C_2}} = 5300$ and $N_{1/e,\mathrm{C_3}} = 3338$.
These decay constants are well beyond the experimentally observed decay constants of 1511 and 1097 (Sec.~\ref{sec:current_best}). These results therefore support the existence of additional noise sources that limit the memory performance in this regime of $\Delta\omega$.\\
Future experiments may trade off repumping duration and off-resonant excitation to obtain improved NV initialization fidelities. The techniques presented in this section allow for the certification of NV initialization errors at the $10^{-5}$ level.
Off-resonant excitation might be further tackled by utilizing DC Stark tuning to steer the excited state level structure and excitation frequency spectrum of the NV electron spin~\cite{tamarat_stark_2006}.\\
\section{Discussion and outlook}
\label{sec:outlook}

In summary, we performed a detailed study of decoherence mechanisms within a quantum network node consisting of an NV centre as optical interface and surrounding $\C$ nuclear spins as quantum memories.
Instead of the earlier suspected stochastic NV reinitialization process, we found that control infidelities and quasi-static noise constitute the major contributors to nuclear-spin decoherence for spins with coupling strengths below $2\pi\times 400\,\mathrm{kHz}$.
These insights lead to the demonstration of a 19-fold improved memory robustness which is still not limited by the inherently stochastic electron reinitialization process.\\

\begin{figure}[tb!]
	\centering
	\includegraphics[width=\columnwidth]{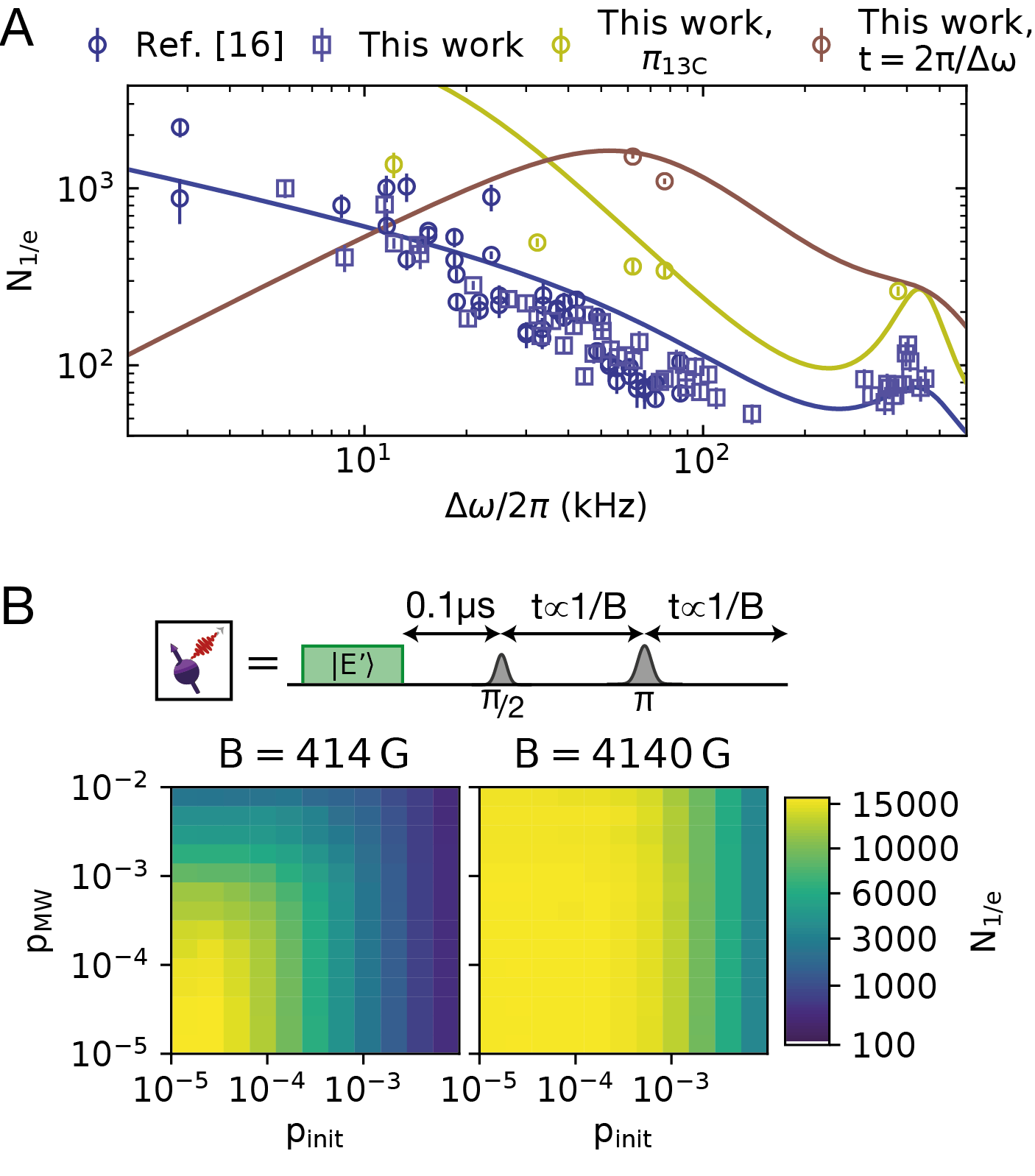}
	\caption{(\textbf{A}) Comparison between experimentally obtained decay constants (see legend) as a function of coupling strength $\Delta\omega$ and a phenomenological model (solid lines). (\textbf{B}) Monte-Carlo simulation of a nuclear spin memory with $\Delta\omega = 26\,\mathrm{kHz}$. We vary the error probabilities for NV electron spin rotations $p_\mathrm{MW}$ and unfaithful initialization $p_\mathrm{init}$ and estimate the $1/e$-decay constant $N_\mathrm{1/e}$ of the memory for two magnetic fields $B$.  We use $\tau=100\,\mathrm{ns}$ resulting in an optimal decay constant of $\sim15\times 10^3$ which is limited by the stochastic electron reinitialization process~\cite{blok_towards_2015}. Our simulation suggests that larger errors are tolerable in a regime of higher magnetic field as the sequence duration of each entangling attempt may be shortened which in turn reduces the accumulated nuclear-spin phase upon error. For a magnetic field of $4.14\,\mathrm{kG}$ the time between microwave pulses becomes $256\,\mathrm{ns} = 2\pi/\omega_0$ while the required microwave frequency for electron spin manipulation becomes $8.7\,\mathrm{GHz}$. Both values are obtainable with state-of-the-art technology.}
	\label{fig:Fig5}
\end{figure}

In Fig.~\ref{fig:Fig5}A we plot the decay constant data of Ref.~\cite{reiserer_robust_2016} (purple circles) as a function of $\Delta\omega$. 
The data set for all seven nuclear spins that are available around the NV used in this work, including two-nuclear-spin subspace configurations, is shown in addition (purple squares) and found to be consistent with the data of Ref.~\cite{reiserer_robust_2016}.
We compare all measured decay constants to a phenomenological model that comprises the repumping process and the control infidelities $p_\mathrm{init}=1.2\cdot10^{-3}$ and $p_\mathrm{MW} = 8\cdot10^{-3}$ (estimated from electron spin measurements). 
The quasi-static noise is hypothesized to originate from laser intensity fluctuations at the electron-spin position and therefore dephasing noise with a phenomenological $\Delta\omega$-dependence is incorporated.
We further show the decay constants when interleaving a nuclear $\pi$-rotation (yellow, see Figs.~\ref{fig:Fig1},\ref{fig:Fig2}) and when utilizing the phase-matching sequences of Sec.~\ref{sec:current_best} (brown) while the respective modelled decay constants are shown as solid lines of the same colour.
 We find reasonable agreement between our model and experimental data when incorporating the Gaussian nuclear coherence decay with $T_\mathrm{2,Hahn} = 60\,\mathrm{ms}$ which stems from the fluctuating spin bath. \\

Our results point towards straightforward improvements to show that memory performance can be greatly increased. 
The spurious phases acquired by nuclear spins during entangling sequences scale with $\Delta\omega$ and the entangling sequence duration. 
This duration is dictated by the Larmor period $\tau_L = 2\pi/\omega_0 \propto 1/B$ of the nuclear spin bath: decoupling the electron with an inter-pulse delay equal to $\tau_L$ preserves the electron spin coherence upon entangling success and induces minimal depolarization noise on the nuclear spin memories by avoiding undesired off-axis rotations~\cite{reiserer_robust_2016}.
It is therefore desirable to increase the magnetic field and in turn shorten the critical phase-sensitive parts of the entangling sequence.\\

To underpin this hypothesis we perform a Monte-Carlo simulation of our experiment for two differing magnetic fields. 
Our Monte-Carlo simulation builds on the simple model of Ref.~\cite{blok_towards_2015} by exponentially distributing electron reinitialization times $t_{\ket{0}}$.
We choose the mean of this distribution as $\tau=100\,\mathrm{ns}$, in accordance with the best measured $\tau$ for nuclear spins $\mathrm{C_2}$ and $\mathrm{C_3}$ in Sec.~\ref{sec:singlet}.
We additionally keep track of the electron state during each entangling attempt and randomly draw microwave and initialization failure events with probabilities $p_\mathrm{MW}$ and $p_\mathrm{init}$. \\

Figure~\ref{fig:Fig5}B presents the simulated decay constants for a nuclear spin with $\Delta\omega= 2\pi\times 26\,\mathrm{kHz}$.
We chose this value for $\Delta\omega$ because nuclear spins in this coupling regime still have an experimental track record of high addressability~\cite{cramer_repeated_2016,reiserer_robust_2016,kalb_experimental_2016,kalb_entanglement_2017} while we expect them to outperform the memories in this work ($\Delta\omega>2\pi\times60\,\mathrm{kHz}$). 
Our simulation predicts a steep increase in memory robustness for higher magnetic fields at already demonstrated error rates while a memory with the chosen $\Delta\omega$ at current magnetic fields ($B=414\,\mathrm{G}$) would only unfold its full potential at the highest-level of experimental control achieved to date~\cite{gaebler_high-fidelity_2016,harty_high-fidelity_2014}.
Moreover operating at an elevated magnetic field will further suppress depolarizing noise originating from the perpendicular hyperfine coupling $A_\perp$~\cite{reiserer_robust_2016}.\\

Decreasing the entangling duration to enhance memory robustness trivially works for short distances between network nodes such that successful entanglement generation events can be heralded within hundreds of nanoseconds. 
For larger distances between network nodes one may achieve similar results by employing multiple inversion pulses on the electron spin, potentially in conjunction with pulse error cancelling dynamical decoupling sequences~\cite{lange_universal_2010,abobeih_one-second_2018}. 
One can additionally mitigate faulty initialization into the electron spin triplet state that is not part of the qubit subspace by using a dedicated continuous light field that couples this unused state to the optically excited state.\\

In conclusion, we have shown that quantum superposition states on weakly-coupled nuclear spins ($\Delta\omega<2\pi \times 80\,\mathrm{kHz}$) are robust against a large number of entangling attempts on the NV electron spin (>1000) and identified quasi-static noise and microwave control errors as the previously limiting factors. 
The exact composition of the currently limiting noise sources for the memory performance remains elusive, but could be further investigated by employing fast laser-intensity stabilization techniques in combination with more nuclear spin inversion rotations and increased magnetic fields.
The obtained results are readily generalized to other solid-state defects and quantum information processing platforms that utilize always-on interactions~\cite{kolesov_optical_2012,christle_isolated_2015,zhong_optically_2015,sukachev_silicon-vacancy_2017,iwasaki_tin-vacancy_2017,morse_photonic_2017,lim_coherent_2017}. 
The majority of nuclear spins surrounding NVs in diamond of natural isotopic composition therefore have a robustness which is comparable to the inverse success probability of generating entanglement at close distances ($10^{-3}$-$10^{-4}$)~\cite{humphreys_deterministic_2017}.
These results further unlock the potential of NV centres as highly-coherent multi-qubit network nodes and may lead to the proof-of-principle demonstrations of an NV-based quantum repeater~\cite{briegel_quantum_1998,rozpedek_realistic_2017} and distributed quantum computation~\cite{nickerson_freely_2014}.\\

\begin{acknowledgements}
	$\quad$ \newline
	We thank M.~Markham and D.~J.~Twitchen from Element Six Inc. for the diamond substrate, M. Pompili for experimental assistance and T.~H.~Taminiau for helpful discussions. We acknowledge support from the Netherlands Organisation for Scientific Research (NWO) through a VICI grant (RH), and the European Research Council through a Synergy Grant (RH).
\end{acknowledgements}

\bibliography{bib}

\begin{thebibliography}{45}%
\makeatletter
\providecommand \@ifxundefined [1]{%
 \@ifx{#1\undefined}
}%
\providecommand \@ifnum [1]{%
 \ifnum #1\expandafter \@firstoftwo
 \else \expandafter \@secondoftwo
 \fi
}%
\providecommand \@ifx [1]{%
 \ifx #1\expandafter \@firstoftwo
 \else \expandafter \@secondoftwo
 \fi
}%
\providecommand \natexlab [1]{#1}%
\providecommand \enquote  [1]{``#1''}%
\providecommand \bibnamefont  [1]{#1}%
\providecommand \bibfnamefont [1]{#1}%
\providecommand \citenamefont [1]{#1}%
\providecommand \href@noop [0]{\@secondoftwo}%
\providecommand \href [0]{\begingroup \@sanitize@url \@href}%
\providecommand \@href[1]{\@@startlink{#1}\@@href}%
\providecommand \@@href[1]{\endgroup#1\@@endlink}%
\providecommand \@sanitize@url [0]{\catcode `\\12\catcode `\$12\catcode
  `\&12\catcode `\#12\catcode `\^12\catcode `\_12\catcode `\%12\relax}%
\providecommand \@@startlink[1]{}%
\providecommand \@@endlink[0]{}%
\providecommand \url  [0]{\begingroup\@sanitize@url \@url }%
\providecommand \@url [1]{\endgroup\@href {#1}{\urlprefix }}%
\providecommand \urlprefix  [0]{URL }%
\providecommand \Eprint [0]{\href }%
\providecommand \doibase [0]{http://dx.doi.org/}%
\providecommand \selectlanguage [0]{\@gobble}%
\providecommand \bibinfo  [0]{\@secondoftwo}%
\providecommand \bibfield  [0]{\@secondoftwo}%
\providecommand \translation [1]{[#1]}%
\providecommand \BibitemOpen [0]{}%
\providecommand \bibitemStop [0]{}%
\providecommand \bibitemNoStop [0]{.\EOS\space}%
\providecommand \EOS [0]{\spacefactor3000\relax}%
\providecommand \BibitemShut  [1]{\csname bibitem#1\endcsname}%
\let\auto@bib@innerbib\@empty
\bibitem [{\citenamefont {Kimble}(2008)}]{kimble_quantum_2008}%
  \BibitemOpen
  \bibfield  {author} {\bibinfo {author} {\bibfnamefont {H.~J.}\ \bibnamefont
  {Kimble}},\ }\href {\doibase 10.1038/nature07127} {\bibfield  {journal}
  {\bibinfo  {journal} {Nature}\ }\textbf {\bibinfo {volume} {453}},\ \bibinfo
  {pages} {1023} (\bibinfo {year} {2008})}\BibitemShut {NoStop}%
\bibitem [{\citenamefont {Moehring}\ \emph {et~al.}(2007)\citenamefont
  {Moehring}, \citenamefont {Maunz}, \citenamefont {Olmschenk}, \citenamefont
  {Younge}, \citenamefont {Matsukevich}, \citenamefont {Duan},\ and\
  \citenamefont {Monroe}}]{moehring_entanglement_2007}%
  \BibitemOpen
  \bibfield  {author} {\bibinfo {author} {\bibfnamefont {D.~L.}\ \bibnamefont
  {Moehring}}, \bibinfo {author} {\bibfnamefont {P.}~\bibnamefont {Maunz}},
  \bibinfo {author} {\bibfnamefont {S.}~\bibnamefont {Olmschenk}}, \bibinfo
  {author} {\bibfnamefont {K.~C.}\ \bibnamefont {Younge}}, \bibinfo {author}
  {\bibfnamefont {D.~N.}\ \bibnamefont {Matsukevich}}, \bibinfo {author}
  {\bibfnamefont {L.-M.}\ \bibnamefont {Duan}}, \ and\ \bibinfo {author}
  {\bibfnamefont {C.}~\bibnamefont {Monroe}},\ }\href {\doibase
  10.1038/nature06118} {\bibfield  {journal} {\bibinfo  {journal} {Nature}\
  }\textbf {\bibinfo {volume} {449}},\ \bibinfo {pages} {68} (\bibinfo {year}
  {2007})}\BibitemShut {NoStop}%
\bibitem [{\citenamefont {Ritter}\ \emph {et~al.}(2012)\citenamefont {Ritter},
  \citenamefont {N{\"o}lleke}, \citenamefont {Hahn}, \citenamefont {Reiserer},
  \citenamefont {Neuzner}, \citenamefont {Uphoff}, \citenamefont {M{\"u}cke},
  \citenamefont {Figueroa}, \citenamefont {Bochmann},\ and\ \citenamefont
  {Rempe}}]{ritter_elementary_2012}%
  \BibitemOpen
  \bibfield  {author} {\bibinfo {author} {\bibfnamefont {S.}~\bibnamefont
  {Ritter}}, \bibinfo {author} {\bibfnamefont {C.}~\bibnamefont {N{\"o}lleke}},
  \bibinfo {author} {\bibfnamefont {C.}~\bibnamefont {Hahn}}, \bibinfo {author}
  {\bibfnamefont {A.}~\bibnamefont {Reiserer}}, \bibinfo {author}
  {\bibfnamefont {A.}~\bibnamefont {Neuzner}}, \bibinfo {author} {\bibfnamefont
  {M.}~\bibnamefont {Uphoff}}, \bibinfo {author} {\bibfnamefont
  {M.}~\bibnamefont {M{\"u}cke}}, \bibinfo {author} {\bibfnamefont
  {E.}~\bibnamefont {Figueroa}}, \bibinfo {author} {\bibfnamefont
  {J.}~\bibnamefont {Bochmann}}, \ and\ \bibinfo {author} {\bibfnamefont
  {G.}~\bibnamefont {Rempe}},\ }\href {\doibase 10.1038/nature11023} {\bibfield
   {journal} {\bibinfo  {journal} {Nature}\ }\textbf {\bibinfo {volume}
  {484}},\ \bibinfo {pages} {195} (\bibinfo {year} {2012})}\BibitemShut
  {NoStop}%
\bibitem [{\citenamefont {Hofmann}\ \emph {et~al.}(2012)\citenamefont
  {Hofmann}, \citenamefont {Krug}, \citenamefont {Ortegel}, \citenamefont
  {G{\'e}rard}, \citenamefont {Weber}, \citenamefont {Rosenfeld},\ and\
  \citenamefont {Weinfurter}}]{hofmann_heralded_2012}%
  \BibitemOpen
  \bibfield  {author} {\bibinfo {author} {\bibfnamefont {J.}~\bibnamefont
  {Hofmann}}, \bibinfo {author} {\bibfnamefont {M.}~\bibnamefont {Krug}},
  \bibinfo {author} {\bibfnamefont {N.}~\bibnamefont {Ortegel}}, \bibinfo
  {author} {\bibfnamefont {L.}~\bibnamefont {G{\'e}rard}}, \bibinfo {author}
  {\bibfnamefont {M.}~\bibnamefont {Weber}}, \bibinfo {author} {\bibfnamefont
  {W.}~\bibnamefont {Rosenfeld}}, \ and\ \bibinfo {author} {\bibfnamefont
  {H.}~\bibnamefont {Weinfurter}},\ }\href {\doibase 10.1126/science.1221856}
  {\bibfield  {journal} {\bibinfo  {journal} {Science}\ }\textbf {\bibinfo
  {volume} {337}},\ \bibinfo {pages} {72} (\bibinfo {year} {2012})}\BibitemShut
  {NoStop}%
\bibitem [{\citenamefont {Bernien}\ \emph {et~al.}(2013)\citenamefont
  {Bernien}, \citenamefont {Hensen}, \citenamefont {Pfaff}, \citenamefont
  {Koolstra}, \citenamefont {Blok}, \citenamefont {Robledo}, \citenamefont
  {Taminiau}, \citenamefont {Markham}, \citenamefont {Twitchen}, \citenamefont
  {Childress},\ and\ \citenamefont {Hanson}}]{bernien_heralded_2013}%
  \BibitemOpen
  \bibfield  {author} {\bibinfo {author} {\bibfnamefont {H.}~\bibnamefont
  {Bernien}}, \bibinfo {author} {\bibfnamefont {B.}~\bibnamefont {Hensen}},
  \bibinfo {author} {\bibfnamefont {W.}~\bibnamefont {Pfaff}}, \bibinfo
  {author} {\bibfnamefont {G.}~\bibnamefont {Koolstra}}, \bibinfo {author}
  {\bibfnamefont {M.~S.}\ \bibnamefont {Blok}}, \bibinfo {author}
  {\bibfnamefont {L.}~\bibnamefont {Robledo}}, \bibinfo {author} {\bibfnamefont
  {T.~H.}\ \bibnamefont {Taminiau}}, \bibinfo {author} {\bibfnamefont
  {M.}~\bibnamefont {Markham}}, \bibinfo {author} {\bibfnamefont {D.~J.}\
  \bibnamefont {Twitchen}}, \bibinfo {author} {\bibfnamefont {L.}~\bibnamefont
  {Childress}}, \ and\ \bibinfo {author} {\bibfnamefont {R.}~\bibnamefont
  {Hanson}},\ }\href {\doibase 10.1038/nature12016} {\bibfield  {journal}
  {\bibinfo  {journal} {Nature}\ }\textbf {\bibinfo {volume} {497}},\ \bibinfo
  {pages} {86} (\bibinfo {year} {2013})}\BibitemShut {NoStop}%
\bibitem [{\citenamefont {Northup}\ and\ \citenamefont
  {Blatt}(2014)}]{northup_quantum_2014}%
  \BibitemOpen
  \bibfield  {author} {\bibinfo {author} {\bibfnamefont {T.~E.}\ \bibnamefont
  {Northup}}\ and\ \bibinfo {author} {\bibfnamefont {R.}~\bibnamefont
  {Blatt}},\ }\href {\doibase 10.1038/nphoton.2014.53} {\bibfield  {journal}
  {\bibinfo  {journal} {Nat. Photonics}\ }\textbf {\bibinfo {volume} {8}},\
  \bibinfo {pages} {356} (\bibinfo {year} {2014})}\BibitemShut {NoStop}%
\bibitem [{\citenamefont {Narla}\ \emph {et~al.}(2016)\citenamefont {Narla},
  \citenamefont {Shankar}, \citenamefont {Hatridge}, \citenamefont {Leghtas},
  \citenamefont {Sliwa}, \citenamefont {Zalys-Geller}, \citenamefont
  {Mundhada}, \citenamefont {Pfaff}, \citenamefont {Frunzio}, \citenamefont
  {Schoelkopf},\ and\ \citenamefont {Devoret}}]{narla_robust_2016}%
  \BibitemOpen
  \bibfield  {author} {\bibinfo {author} {\bibfnamefont {A.}~\bibnamefont
  {Narla}}, \bibinfo {author} {\bibfnamefont {S.}~\bibnamefont {Shankar}},
  \bibinfo {author} {\bibfnamefont {M.}~\bibnamefont {Hatridge}}, \bibinfo
  {author} {\bibfnamefont {Z.}~\bibnamefont {Leghtas}}, \bibinfo {author}
  {\bibfnamefont {K.~M.}\ \bibnamefont {Sliwa}}, \bibinfo {author}
  {\bibfnamefont {E.}~\bibnamefont {Zalys-Geller}}, \bibinfo {author}
  {\bibfnamefont {S.~O.}\ \bibnamefont {Mundhada}}, \bibinfo {author}
  {\bibfnamefont {W.}~\bibnamefont {Pfaff}}, \bibinfo {author} {\bibfnamefont
  {L.}~\bibnamefont {Frunzio}}, \bibinfo {author} {\bibfnamefont {R.~J.}\
  \bibnamefont {Schoelkopf}}, \ and\ \bibinfo {author} {\bibfnamefont {M.~H.}\
  \bibnamefont {Devoret}},\ }\href {\doibase 10.1103/PhysRevX.6.031036}
  {\bibfield  {journal} {\bibinfo  {journal} {Phys. Rev. X}\ }\textbf {\bibinfo
  {volume} {6}},\ \bibinfo {pages} {031036} (\bibinfo {year}
  {2016})}\BibitemShut {NoStop}%
\bibitem [{\citenamefont {Stockill}\ \emph {et~al.}(2017)\citenamefont
  {Stockill}, \citenamefont {Stanley}, \citenamefont {Huthmacher},
  \citenamefont {Clarke}, \citenamefont {Hugues}, \citenamefont {Miller},
  \citenamefont {Matthiesen}, \citenamefont {Le~Gall},\ and\ \citenamefont
  {Atat{\"u}re}}]{stockill_phase-tuned_2017}%
  \BibitemOpen
  \bibfield  {author} {\bibinfo {author} {\bibfnamefont {R.}~\bibnamefont
  {Stockill}}, \bibinfo {author} {\bibfnamefont {M.~J.}\ \bibnamefont
  {Stanley}}, \bibinfo {author} {\bibfnamefont {L.}~\bibnamefont {Huthmacher}},
  \bibinfo {author} {\bibfnamefont {E.}~\bibnamefont {Clarke}}, \bibinfo
  {author} {\bibfnamefont {M.}~\bibnamefont {Hugues}}, \bibinfo {author}
  {\bibfnamefont {A.~J.}\ \bibnamefont {Miller}}, \bibinfo {author}
  {\bibfnamefont {C.}~\bibnamefont {Matthiesen}}, \bibinfo {author}
  {\bibfnamefont {C.}~\bibnamefont {Le~Gall}}, \ and\ \bibinfo {author}
  {\bibfnamefont {M.}~\bibnamefont {Atat{\"u}re}},\ }\href {\doibase
  10.1103/PhysRevLett.119.010503} {\bibfield  {journal} {\bibinfo  {journal}
  {Phys. Rev. Lett.}\ }\textbf {\bibinfo {volume} {119}},\ \bibinfo {pages}
  {010503} (\bibinfo {year} {2017})}\BibitemShut {NoStop}%
\bibitem [{\citenamefont {Pfaff}\ \emph {et~al.}(2014)\citenamefont {Pfaff},
  \citenamefont {Hensen}, \citenamefont {Bernien}, \citenamefont {van Dam},
  \citenamefont {Blok}, \citenamefont {Taminiau}, \citenamefont {Tiggelman},
  \citenamefont {Schouten}, \citenamefont {Markham}, \citenamefont {Twitchen},\
  and\ \citenamefont {Hanson}}]{pfaff_unconditional_2014}%
  \BibitemOpen
  \bibfield  {author} {\bibinfo {author} {\bibfnamefont {W.}~\bibnamefont
  {Pfaff}}, \bibinfo {author} {\bibfnamefont {B.~J.}\ \bibnamefont {Hensen}},
  \bibinfo {author} {\bibfnamefont {H.}~\bibnamefont {Bernien}}, \bibinfo
  {author} {\bibfnamefont {S.~B.}\ \bibnamefont {van Dam}}, \bibinfo {author}
  {\bibfnamefont {M.~S.}\ \bibnamefont {Blok}}, \bibinfo {author}
  {\bibfnamefont {T.~H.}\ \bibnamefont {Taminiau}}, \bibinfo {author}
  {\bibfnamefont {M.~J.}\ \bibnamefont {Tiggelman}}, \bibinfo {author}
  {\bibfnamefont {R.~N.}\ \bibnamefont {Schouten}}, \bibinfo {author}
  {\bibfnamefont {M.}~\bibnamefont {Markham}}, \bibinfo {author} {\bibfnamefont
  {D.~J.}\ \bibnamefont {Twitchen}}, \ and\ \bibinfo {author} {\bibfnamefont
  {R.}~\bibnamefont {Hanson}},\ }\href {\doibase 10.1126/science.1253512}
  {\bibfield  {journal} {\bibinfo  {journal} {Science}\ }\textbf {\bibinfo
  {volume} {345}},\ \bibinfo {pages} {532} (\bibinfo {year}
  {2014})}\BibitemShut {NoStop}%
\bibitem [{\citenamefont {Hucul}\ \emph {et~al.}(2015)\citenamefont {Hucul},
  \citenamefont {Inlek}, \citenamefont {Vittorini}, \citenamefont {Crocker},
  \citenamefont {Debnath}, \citenamefont {Clark},\ and\ \citenamefont
  {Monroe}}]{hucul_modular_2015}%
  \BibitemOpen
  \bibfield  {author} {\bibinfo {author} {\bibfnamefont {D.}~\bibnamefont
  {Hucul}}, \bibinfo {author} {\bibfnamefont {I.~V.}\ \bibnamefont {Inlek}},
  \bibinfo {author} {\bibfnamefont {G.}~\bibnamefont {Vittorini}}, \bibinfo
  {author} {\bibfnamefont {C.}~\bibnamefont {Crocker}}, \bibinfo {author}
  {\bibfnamefont {S.}~\bibnamefont {Debnath}}, \bibinfo {author} {\bibfnamefont
  {S.~M.}\ \bibnamefont {Clark}}, \ and\ \bibinfo {author} {\bibfnamefont
  {C.}~\bibnamefont {Monroe}},\ }\href {\doibase 10.1038/nphys3150} {\bibfield
  {journal} {\bibinfo  {journal} {Nat. Phys.}\ }\textbf {\bibinfo {volume}
  {11}},\ \bibinfo {pages} {37} (\bibinfo {year} {2015})}\BibitemShut {NoStop}%
\bibitem [{\citenamefont {Kalb}\ \emph {et~al.}(2017)\citenamefont {Kalb},
  \citenamefont {Reiserer}, \citenamefont {Humphreys}, \citenamefont
  {Bakermans}, \citenamefont {Kamerling}, \citenamefont {Nickerson},
  \citenamefont {Benjamin}, \citenamefont {Twitchen}, \citenamefont {Markham},\
  and\ \citenamefont {Hanson}}]{kalb_entanglement_2017}%
  \BibitemOpen
  \bibfield  {author} {\bibinfo {author} {\bibfnamefont {N.}~\bibnamefont
  {Kalb}}, \bibinfo {author} {\bibfnamefont {A.~A.}\ \bibnamefont {Reiserer}},
  \bibinfo {author} {\bibfnamefont {P.~C.}\ \bibnamefont {Humphreys}}, \bibinfo
  {author} {\bibfnamefont {J.~J.~W.}\ \bibnamefont {Bakermans}}, \bibinfo
  {author} {\bibfnamefont {S.~J.}\ \bibnamefont {Kamerling}}, \bibinfo {author}
  {\bibfnamefont {N.~H.}\ \bibnamefont {Nickerson}}, \bibinfo {author}
  {\bibfnamefont {S.~C.}\ \bibnamefont {Benjamin}}, \bibinfo {author}
  {\bibfnamefont {D.~J.}\ \bibnamefont {Twitchen}}, \bibinfo {author}
  {\bibfnamefont {M.}~\bibnamefont {Markham}}, \ and\ \bibinfo {author}
  {\bibfnamefont {R.}~\bibnamefont {Hanson}},\ }\href {\doibase
  10.1126/science.aan0070} {\bibfield  {journal} {\bibinfo  {journal}
  {Science}\ }\textbf {\bibinfo {volume} {356}},\ \bibinfo {pages} {928}
  (\bibinfo {year} {2017})}\BibitemShut {NoStop}%
\bibitem [{\citenamefont {Zhao}\ \emph {et~al.}(2012)\citenamefont {Zhao},
  \citenamefont {Honert}, \citenamefont {Schmid}, \citenamefont {Klas},
  \citenamefont {Isoya}, \citenamefont {Markham}, \citenamefont {Twitchen},
  \citenamefont {Jelezko}, \citenamefont {Liu}, \citenamefont {Fedder},\ and\
  \citenamefont {Wrachtrup}}]{zhao_sensing_2012}%
  \BibitemOpen
  \bibfield  {author} {\bibinfo {author} {\bibfnamefont {N.}~\bibnamefont
  {Zhao}}, \bibinfo {author} {\bibfnamefont {J.}~\bibnamefont {Honert}},
  \bibinfo {author} {\bibfnamefont {B.}~\bibnamefont {Schmid}}, \bibinfo
  {author} {\bibfnamefont {M.}~\bibnamefont {Klas}}, \bibinfo {author}
  {\bibfnamefont {J.}~\bibnamefont {Isoya}}, \bibinfo {author} {\bibfnamefont
  {M.}~\bibnamefont {Markham}}, \bibinfo {author} {\bibfnamefont
  {D.}~\bibnamefont {Twitchen}}, \bibinfo {author} {\bibfnamefont
  {F.}~\bibnamefont {Jelezko}}, \bibinfo {author} {\bibfnamefont {R.-B.}\
  \bibnamefont {Liu}}, \bibinfo {author} {\bibfnamefont {H.}~\bibnamefont
  {Fedder}}, \ and\ \bibinfo {author} {\bibfnamefont {J.}~\bibnamefont
  {Wrachtrup}},\ }\href {\doibase 10.1038/nnano.2012.152} {\bibfield  {journal}
  {\bibinfo  {journal} {Nat. Nanotechnol.}\ }\textbf {\bibinfo {volume} {7}},\
  \bibinfo {pages} {657} (\bibinfo {year} {2012})}\BibitemShut {NoStop}%
\bibitem [{\citenamefont {Kolkowitz}\ \emph {et~al.}(2012)\citenamefont
  {Kolkowitz}, \citenamefont {Unterreithmeier}, \citenamefont {Bennett},\ and\
  \citenamefont {Lukin}}]{kolkowitz_sensing_2012}%
  \BibitemOpen
  \bibfield  {author} {\bibinfo {author} {\bibfnamefont {S.}~\bibnamefont
  {Kolkowitz}}, \bibinfo {author} {\bibfnamefont {Q.~P.}\ \bibnamefont
  {Unterreithmeier}}, \bibinfo {author} {\bibfnamefont {S.~D.}\ \bibnamefont
  {Bennett}}, \ and\ \bibinfo {author} {\bibfnamefont {M.~D.}\ \bibnamefont
  {Lukin}},\ }\href {\doibase 10.1103/PhysRevLett.109.137601} {\bibfield
  {journal} {\bibinfo  {journal} {Phys. Rev. Lett.}\ }\textbf {\bibinfo
  {volume} {109}},\ \bibinfo {pages} {137601} (\bibinfo {year}
  {2012})}\BibitemShut {NoStop}%
\bibitem [{\citenamefont {Taminiau}\ \emph {et~al.}(2014)\citenamefont
  {Taminiau}, \citenamefont {Cramer}, \citenamefont {van~der Sar},
  \citenamefont {Dobrovitski},\ and\ \citenamefont
  {Hanson}}]{taminiau_universal_2014}%
  \BibitemOpen
  \bibfield  {author} {\bibinfo {author} {\bibfnamefont {T.~H.}\ \bibnamefont
  {Taminiau}}, \bibinfo {author} {\bibfnamefont {J.}~\bibnamefont {Cramer}},
  \bibinfo {author} {\bibfnamefont {T.}~\bibnamefont {van~der Sar}}, \bibinfo
  {author} {\bibfnamefont {V.~V.}\ \bibnamefont {Dobrovitski}}, \ and\ \bibinfo
  {author} {\bibfnamefont {R.}~\bibnamefont {Hanson}},\ }\href {\doibase
  10.1038/nnano.2014.2} {\bibfield  {journal} {\bibinfo  {journal} {Nat.
  Nanotechnol.}\ }\textbf {\bibinfo {volume} {9}},\ \bibinfo {pages} {171}
  (\bibinfo {year} {2014})}\BibitemShut {NoStop}%
\bibitem [{\citenamefont {Blok}\ \emph {et~al.}(2015)\citenamefont {Blok},
  \citenamefont {Kalb}, \citenamefont {Reiserer}, \citenamefont {Taminiau},\
  and\ \citenamefont {Hanson}}]{blok_towards_2015}%
  \BibitemOpen
  \bibfield  {author} {\bibinfo {author} {\bibfnamefont {M.~S.}\ \bibnamefont
  {Blok}}, \bibinfo {author} {\bibfnamefont {N.}~\bibnamefont {Kalb}}, \bibinfo
  {author} {\bibfnamefont {A.}~\bibnamefont {Reiserer}}, \bibinfo {author}
  {\bibfnamefont {T.~H.}\ \bibnamefont {Taminiau}}, \ and\ \bibinfo {author}
  {\bibfnamefont {R.}~\bibnamefont {Hanson}},\ }\href {\doibase
  10.1039/C5FD00113G} {\bibfield  {journal} {\bibinfo  {journal} {Faraday
  Discuss.}\ }\textbf {\bibinfo {volume} {184}},\ \bibinfo {pages} {173}
  (\bibinfo {year} {2015})}\BibitemShut {NoStop}%
\bibitem [{\citenamefont {Reiserer}\ \emph {et~al.}(2016)\citenamefont
  {Reiserer}, \citenamefont {Kalb}, \citenamefont {Blok}, \citenamefont {van
  Bemmelen}, \citenamefont {Taminiau}, \citenamefont {Hanson}, \citenamefont
  {Twitchen},\ and\ \citenamefont {Markham}}]{reiserer_robust_2016}%
  \BibitemOpen
  \bibfield  {author} {\bibinfo {author} {\bibfnamefont {A.}~\bibnamefont
  {Reiserer}}, \bibinfo {author} {\bibfnamefont {N.}~\bibnamefont {Kalb}},
  \bibinfo {author} {\bibfnamefont {M.~S.}\ \bibnamefont {Blok}}, \bibinfo
  {author} {\bibfnamefont {K.~J.~M.}\ \bibnamefont {van Bemmelen}}, \bibinfo
  {author} {\bibfnamefont {T.~H.}\ \bibnamefont {Taminiau}}, \bibinfo {author}
  {\bibfnamefont {R.}~\bibnamefont {Hanson}}, \bibinfo {author} {\bibfnamefont
  {D.~J.}\ \bibnamefont {Twitchen}}, \ and\ \bibinfo {author} {\bibfnamefont
  {M.}~\bibnamefont {Markham}},\ }\href {\doibase 10.1103/PhysRevX.6.021040}
  {\bibfield  {journal} {\bibinfo  {journal} {Phys. Rev. X}\ }\textbf {\bibinfo
  {volume} {6}},\ \bibinfo {pages} {021040} (\bibinfo {year}
  {2016})}\BibitemShut {NoStop}%
\bibitem [{\citenamefont {Yeung}\ \emph {et~al.}(2012)\citenamefont {Yeung},
  \citenamefont {Le~Sage}, \citenamefont {Pham}, \citenamefont {Stanwix},\ and\
  \citenamefont {Walsworth}}]{yeung_anti-reflection_2012}%
  \BibitemOpen
  \bibfield  {author} {\bibinfo {author} {\bibfnamefont {T.~K.}\ \bibnamefont
  {Yeung}}, \bibinfo {author} {\bibfnamefont {D.}~\bibnamefont {Le~Sage}},
  \bibinfo {author} {\bibfnamefont {L.~M.}\ \bibnamefont {Pham}}, \bibinfo
  {author} {\bibfnamefont {P.~L.}\ \bibnamefont {Stanwix}}, \ and\ \bibinfo
  {author} {\bibfnamefont {R.~L.}\ \bibnamefont {Walsworth}},\ }\href {\doibase
  10.1063/1.4730401} {\bibfield  {journal} {\bibinfo  {journal} {Appl. Phys.
  Lett.}\ }\textbf {\bibinfo {volume} {100}},\ \bibinfo {pages} {251111}
  (\bibinfo {year} {2012})}\BibitemShut {NoStop}%
\bibitem [{\citenamefont {Robledo}\ \emph
  {et~al.}(2011{\natexlab{a}})\citenamefont {Robledo}, \citenamefont
  {Childress}, \citenamefont {Bernien}, \citenamefont {Hensen}, \citenamefont
  {Alkemade},\ and\ \citenamefont {Hanson}}]{robledo_high-fidelity_2011}%
  \BibitemOpen
  \bibfield  {author} {\bibinfo {author} {\bibfnamefont {L.}~\bibnamefont
  {Robledo}}, \bibinfo {author} {\bibfnamefont {L.}~\bibnamefont {Childress}},
  \bibinfo {author} {\bibfnamefont {H.}~\bibnamefont {Bernien}}, \bibinfo
  {author} {\bibfnamefont {B.}~\bibnamefont {Hensen}}, \bibinfo {author}
  {\bibfnamefont {P.~F.~A.}\ \bibnamefont {Alkemade}}, \ and\ \bibinfo {author}
  {\bibfnamefont {R.}~\bibnamefont {Hanson}},\ }\href {\doibase
  10.1038/nature10401} {\bibfield  {journal} {\bibinfo  {journal} {Nature}\
  }\textbf {\bibinfo {volume} {477}},\ \bibinfo {pages} {574} (\bibinfo {year}
  {2011}{\natexlab{a}})}\BibitemShut {NoStop}%
\bibitem [{\citenamefont {Yang}\ \emph {et~al.}(2016)\citenamefont {Yang},
  \citenamefont {Wang}, \citenamefont {Rao}, \citenamefont {Hien~Tran},
  \citenamefont {Momenzadeh}, \citenamefont {Markham}, \citenamefont
  {Twitchen}, \citenamefont {Wang}, \citenamefont {Yang}, \citenamefont
  {St{\"o}hr}, \citenamefont {Neumann}, \citenamefont {Kosaka},\ and\
  \citenamefont {Wrachtrup}}]{yang_high-fidelity_2016}%
  \BibitemOpen
  \bibfield  {author} {\bibinfo {author} {\bibfnamefont {S.}~\bibnamefont
  {Yang}}, \bibinfo {author} {\bibfnamefont {Y.}~\bibnamefont {Wang}}, \bibinfo
  {author} {\bibfnamefont {D.~D.~B.}\ \bibnamefont {Rao}}, \bibinfo {author}
  {\bibfnamefont {T.}~\bibnamefont {Hien~Tran}}, \bibinfo {author}
  {\bibfnamefont {A.~S.}\ \bibnamefont {Momenzadeh}}, \bibinfo {author}
  {\bibfnamefont {M.}~\bibnamefont {Markham}}, \bibinfo {author} {\bibfnamefont
  {D.~J.}\ \bibnamefont {Twitchen}}, \bibinfo {author} {\bibfnamefont
  {P.}~\bibnamefont {Wang}}, \bibinfo {author} {\bibfnamefont {W.}~\bibnamefont
  {Yang}}, \bibinfo {author} {\bibfnamefont {R.}~\bibnamefont {St{\"o}hr}},
  \bibinfo {author} {\bibfnamefont {P.}~\bibnamefont {Neumann}}, \bibinfo
  {author} {\bibfnamefont {H.}~\bibnamefont {Kosaka}}, \ and\ \bibinfo {author}
  {\bibfnamefont {J.}~\bibnamefont {Wrachtrup}},\ }\href {\doibase
  10.1038/nphoton.2016.103} {\bibfield  {journal} {\bibinfo  {journal} {Nat.
  Photonics}\ }\textbf {\bibinfo {volume} {10}},\ \bibinfo {pages} {507}
  (\bibinfo {year} {2016})}\BibitemShut {NoStop}%
\bibitem [{\citenamefont {Cabrillo}\ \emph {et~al.}(1999)\citenamefont
  {Cabrillo}, \citenamefont {Cirac}, \citenamefont {Garc{\'i}a-Fern{\'a}ndez},\
  and\ \citenamefont {Zoller}}]{cabrillo_creation_1999}%
  \BibitemOpen
  \bibfield  {author} {\bibinfo {author} {\bibfnamefont {C.}~\bibnamefont
  {Cabrillo}}, \bibinfo {author} {\bibfnamefont {J.~I.}\ \bibnamefont {Cirac}},
  \bibinfo {author} {\bibfnamefont {P.}~\bibnamefont
  {Garc{\'i}a-Fern{\'a}ndez}}, \ and\ \bibinfo {author} {\bibfnamefont
  {P.}~\bibnamefont {Zoller}},\ }\href {\doibase 10.1103/PhysRevA.59.1025}
  {\bibfield  {journal} {\bibinfo  {journal} {Phys. Rev. A}\ }\textbf {\bibinfo
  {volume} {59}},\ \bibinfo {pages} {1025} (\bibinfo {year}
  {1999})}\BibitemShut {NoStop}%
\bibitem [{\citenamefont {Barrett}\ and\ \citenamefont
  {Kok}(2005)}]{barrett_efficient_2005}%
  \BibitemOpen
  \bibfield  {author} {\bibinfo {author} {\bibfnamefont {S.~D.}\ \bibnamefont
  {Barrett}}\ and\ \bibinfo {author} {\bibfnamefont {P.}~\bibnamefont {Kok}},\
  }\href {\doibase 10.1103/PhysRevA.71.060310} {\bibfield  {journal} {\bibinfo
  {journal} {Phys. Rev. A}\ }\textbf {\bibinfo {volume} {71}},\ \bibinfo
  {pages} {060310} (\bibinfo {year} {2005})}\BibitemShut {NoStop}%
\bibitem [{\citenamefont {Campbell}\ and\ \citenamefont
  {Benjamin}(2008)}]{campbell_measurement-based_2008}%
  \BibitemOpen
  \bibfield  {author} {\bibinfo {author} {\bibfnamefont {E.~T.}\ \bibnamefont
  {Campbell}}\ and\ \bibinfo {author} {\bibfnamefont {S.~C.}\ \bibnamefont
  {Benjamin}},\ }\href {\doibase 10.1103/PhysRevLett.101.130502} {\bibfield
  {journal} {\bibinfo  {journal} {Phys. Rev. Lett.}\ }\textbf {\bibinfo
  {volume} {101}},\ \bibinfo {pages} {130502} (\bibinfo {year}
  {2008})}\BibitemShut {NoStop}%
\bibitem [{\citenamefont {Doherty}\ \emph {et~al.}(2013)\citenamefont
  {Doherty}, \citenamefont {Manson}, \citenamefont {Delaney}, \citenamefont
  {Jelezko}, \citenamefont {Wrachtrup},\ and\ \citenamefont
  {Hollenberg}}]{doherty_nitrogen-vacancy_2013}%
  \BibitemOpen
  \bibfield  {author} {\bibinfo {author} {\bibfnamefont {M.~W.}\ \bibnamefont
  {Doherty}}, \bibinfo {author} {\bibfnamefont {N.~B.}\ \bibnamefont {Manson}},
  \bibinfo {author} {\bibfnamefont {P.}~\bibnamefont {Delaney}}, \bibinfo
  {author} {\bibfnamefont {F.}~\bibnamefont {Jelezko}}, \bibinfo {author}
  {\bibfnamefont {J.}~\bibnamefont {Wrachtrup}}, \ and\ \bibinfo {author}
  {\bibfnamefont {L.~C.~L.}\ \bibnamefont {Hollenberg}},\ }\href {\doibase
  10.1016/j.physrep.2013.02.001} {\bibfield  {journal} {\bibinfo  {journal}
  {Phys. Rep.}\ }\textbf {\bibinfo {volume} {528}},\ \bibinfo {pages} {1}
  (\bibinfo {year} {2013})}\BibitemShut {NoStop}%
\bibitem [{\citenamefont {Tamarat}\ \emph {et~al.}(2006)\citenamefont
  {Tamarat}, \citenamefont {Gaebel}, \citenamefont {Rabeau}, \citenamefont
  {Khan}, \citenamefont {Greentree}, \citenamefont {Wilson}, \citenamefont
  {Hollenberg}, \citenamefont {Prawer}, \citenamefont {Hemmer}, \citenamefont
  {Jelezko},\ and\ \citenamefont {Wrachtrup}}]{tamarat_stark_2006}%
  \BibitemOpen
  \bibfield  {author} {\bibinfo {author} {\bibfnamefont {P.}~\bibnamefont
  {Tamarat}}, \bibinfo {author} {\bibfnamefont {T.}~\bibnamefont {Gaebel}},
  \bibinfo {author} {\bibfnamefont {J.~R.}\ \bibnamefont {Rabeau}}, \bibinfo
  {author} {\bibfnamefont {M.}~\bibnamefont {Khan}}, \bibinfo {author}
  {\bibfnamefont {A.~D.}\ \bibnamefont {Greentree}}, \bibinfo {author}
  {\bibfnamefont {H.}~\bibnamefont {Wilson}}, \bibinfo {author} {\bibfnamefont
  {L.~C.~L.}\ \bibnamefont {Hollenberg}}, \bibinfo {author} {\bibfnamefont
  {S.}~\bibnamefont {Prawer}}, \bibinfo {author} {\bibfnamefont
  {P.}~\bibnamefont {Hemmer}}, \bibinfo {author} {\bibfnamefont
  {F.}~\bibnamefont {Jelezko}}, \ and\ \bibinfo {author} {\bibfnamefont
  {J.}~\bibnamefont {Wrachtrup}},\ }\href {\doibase
  10.1103/PhysRevLett.97.083002} {\bibfield  {journal} {\bibinfo  {journal}
  {Phys. Rev. Lett.}\ }\textbf {\bibinfo {volume} {97}},\ \bibinfo {pages}
  {083002} (\bibinfo {year} {2006})}\BibitemShut {NoStop}%
\bibitem [{\citenamefont {Robledo}\ \emph
  {et~al.}(2011{\natexlab{b}})\citenamefont {Robledo}, \citenamefont {Bernien},
  \citenamefont {van~der Sar},\ and\ \citenamefont
  {Hanson}}]{robledo_spin_2011}%
  \BibitemOpen
  \bibfield  {author} {\bibinfo {author} {\bibfnamefont {L.}~\bibnamefont
  {Robledo}}, \bibinfo {author} {\bibfnamefont {H.}~\bibnamefont {Bernien}},
  \bibinfo {author} {\bibfnamefont {T.}~\bibnamefont {van~der Sar}}, \ and\
  \bibinfo {author} {\bibfnamefont {R.}~\bibnamefont {Hanson}},\ }\href
  {\doibase 10.1088/1367-2630/13/2/025013} {\bibfield  {journal} {\bibinfo
  {journal} {New J. Phys.}\ }\textbf {\bibinfo {volume} {13}},\ \bibinfo
  {pages} {025013} (\bibinfo {year} {2011}{\natexlab{b}})}\BibitemShut
  {NoStop}%
\bibitem [{\citenamefont {Acosta}\ \emph {et~al.}(2010)\citenamefont {Acosta},
  \citenamefont {Jarmola}, \citenamefont {Bauch},\ and\ \citenamefont
  {Budker}}]{acosta_optical_2010}%
  \BibitemOpen
  \bibfield  {author} {\bibinfo {author} {\bibfnamefont {V.~M.}\ \bibnamefont
  {Acosta}}, \bibinfo {author} {\bibfnamefont {A.}~\bibnamefont {Jarmola}},
  \bibinfo {author} {\bibfnamefont {E.}~\bibnamefont {Bauch}}, \ and\ \bibinfo
  {author} {\bibfnamefont {D.}~\bibnamefont {Budker}},\ }\href {\doibase
  10.1103/PhysRevB.82.201202} {\bibfield  {journal} {\bibinfo  {journal} {Phys.
  Rev. B}\ }\textbf {\bibinfo {volume} {82}},\ \bibinfo {pages} {201202}
  (\bibinfo {year} {2010})}\BibitemShut {NoStop}%
\bibitem [{\citenamefont {Goldman}\ \emph {et~al.}(2015)\citenamefont
  {Goldman}, \citenamefont {Sipahigil}, \citenamefont {Doherty}, \citenamefont
  {Yao}, \citenamefont {Bennett}, \citenamefont {Markham}, \citenamefont
  {Twitchen}, \citenamefont {Manson}, \citenamefont {Kubanek},\ and\
  \citenamefont {Lukin}}]{goldman_phonon-induced_2015-1}%
  \BibitemOpen
  \bibfield  {author} {\bibinfo {author} {\bibfnamefont {M.~L.}\ \bibnamefont
  {Goldman}}, \bibinfo {author} {\bibfnamefont {A.}~\bibnamefont {Sipahigil}},
  \bibinfo {author} {\bibfnamefont {M.~W.}\ \bibnamefont {Doherty}}, \bibinfo
  {author} {\bibfnamefont {N.~Y.}\ \bibnamefont {Yao}}, \bibinfo {author}
  {\bibfnamefont {S.~D.}\ \bibnamefont {Bennett}}, \bibinfo {author}
  {\bibfnamefont {M.}~\bibnamefont {Markham}}, \bibinfo {author} {\bibfnamefont
  {D.~J.}\ \bibnamefont {Twitchen}}, \bibinfo {author} {\bibfnamefont {N.~B.}\
  \bibnamefont {Manson}}, \bibinfo {author} {\bibfnamefont {A.}~\bibnamefont
  {Kubanek}}, \ and\ \bibinfo {author} {\bibfnamefont {M.~D.}\ \bibnamefont
  {Lukin}},\ }\href {\doibase 10.1103/PhysRevLett.114.145502} {\bibfield
  {journal} {\bibinfo  {journal} {Phys. Rev. Lett.}\ }\textbf {\bibinfo
  {volume} {114}},\ \bibinfo {pages} {145502} (\bibinfo {year}
  {2015})}\BibitemShut {NoStop}%
\bibitem [{\citenamefont {Humphreys}\ \emph {et~al.}()\citenamefont
  {Humphreys}, \citenamefont {Kalb}, \citenamefont {Morits}, \citenamefont
  {Schouten}, \citenamefont {Vermeulen}, \citenamefont {Twitchen},
  \citenamefont {Markham},\ and\ \citenamefont
  {Hanson}}]{humphreys_deterministic_2017}%
  \BibitemOpen
  \bibfield  {author} {\bibinfo {author} {\bibfnamefont {P.~C.}\ \bibnamefont
  {Humphreys}}, \bibinfo {author} {\bibfnamefont {N.}~\bibnamefont {Kalb}},
  \bibinfo {author} {\bibfnamefont {J.~P.~J.}\ \bibnamefont {Morits}}, \bibinfo
  {author} {\bibfnamefont {R.~N.}\ \bibnamefont {Schouten}}, \bibinfo {author}
  {\bibfnamefont {R.~F.~L.}\ \bibnamefont {Vermeulen}}, \bibinfo {author}
  {\bibfnamefont {D.~J.}\ \bibnamefont {Twitchen}}, \bibinfo {author}
  {\bibfnamefont {M.}~\bibnamefont {Markham}}, \ and\ \bibinfo {author}
  {\bibfnamefont {R.}~\bibnamefont {Hanson}},\ }\href
  {http://arxiv.org/abs/1712.07567} {\bibinfo  {journal} {arXiv:1712.07567}\
  }\BibitemShut {NoStop}%
\bibitem [{\citenamefont {Felton}\ \emph {et~al.}(2009)\citenamefont {Felton},
  \citenamefont {Edmonds}, \citenamefont {Newton}, \citenamefont {Martineau},
  \citenamefont {Fisher}, \citenamefont {Twitchen},\ and\ \citenamefont
  {Baker}}]{felton_hyperfine_2009}%
  \BibitemOpen
\bibfield  {journal} {  }\bibfield  {author} {\bibinfo {author} {\bibfnamefont
  {S.}~\bibnamefont {Felton}}, \bibinfo {author} {\bibfnamefont {A.~M.}\
  \bibnamefont {Edmonds}}, \bibinfo {author} {\bibfnamefont {M.~E.}\
  \bibnamefont {Newton}}, \bibinfo {author} {\bibfnamefont {P.~M.}\
  \bibnamefont {Martineau}}, \bibinfo {author} {\bibfnamefont {D.}~\bibnamefont
  {Fisher}}, \bibinfo {author} {\bibfnamefont {D.~J.}\ \bibnamefont
  {Twitchen}}, \ and\ \bibinfo {author} {\bibfnamefont {J.~M.}\ \bibnamefont
  {Baker}},\ }\href {\doibase 10.1103/PhysRevB.79.075203} {\bibfield  {journal}
  {\bibinfo  {journal} {Phys. Rev. B}\ }\textbf {\bibinfo {volume} {79}},\
  \bibinfo {pages} {075203} (\bibinfo {year} {2009})}\BibitemShut {NoStop}%
\bibitem [{\citenamefont {Zhong}\ \emph {et~al.}(2015)\citenamefont {Zhong},
  \citenamefont {Hedges}, \citenamefont {Ahlefeldt}, \citenamefont
  {Bartholomew}, \citenamefont {Beavan}, \citenamefont {Wittig}, \citenamefont
  {Longdell},\ and\ \citenamefont {Sellars}}]{zhong_optically_2015}%
  \BibitemOpen
  \bibfield  {author} {\bibinfo {author} {\bibfnamefont {M.}~\bibnamefont
  {Zhong}}, \bibinfo {author} {\bibfnamefont {M.~P.}\ \bibnamefont {Hedges}},
  \bibinfo {author} {\bibfnamefont {R.~L.}\ \bibnamefont {Ahlefeldt}}, \bibinfo
  {author} {\bibfnamefont {J.~G.}\ \bibnamefont {Bartholomew}}, \bibinfo
  {author} {\bibfnamefont {S.~E.}\ \bibnamefont {Beavan}}, \bibinfo {author}
  {\bibfnamefont {S.~M.}\ \bibnamefont {Wittig}}, \bibinfo {author}
  {\bibfnamefont {J.~J.}\ \bibnamefont {Longdell}}, \ and\ \bibinfo {author}
  {\bibfnamefont {M.~J.}\ \bibnamefont {Sellars}},\ }\href {\doibase
  10.1038/nature14025} {\bibfield  {journal} {\bibinfo  {journal} {Nature}\
  }\textbf {\bibinfo {volume} {517}},\ \bibinfo {pages} {177} (\bibinfo {year}
  {2015})}\BibitemShut {NoStop}%
\bibitem [{\citenamefont {Cramer}\ \emph {et~al.}(2016)\citenamefont {Cramer},
  \citenamefont {Kalb}, \citenamefont {Rol}, \citenamefont {Hensen},
  \citenamefont {Blok}, \citenamefont {Markham}, \citenamefont {Twitchen},
  \citenamefont {Hanson},\ and\ \citenamefont
  {Taminiau}}]{cramer_repeated_2016}%
  \BibitemOpen
  \bibfield  {author} {\bibinfo {author} {\bibfnamefont {J.}~\bibnamefont
  {Cramer}}, \bibinfo {author} {\bibfnamefont {N.}~\bibnamefont {Kalb}},
  \bibinfo {author} {\bibfnamefont {M.~A.}\ \bibnamefont {Rol}}, \bibinfo
  {author} {\bibfnamefont {B.}~\bibnamefont {Hensen}}, \bibinfo {author}
  {\bibfnamefont {M.~S.}\ \bibnamefont {Blok}}, \bibinfo {author}
  {\bibfnamefont {M.}~\bibnamefont {Markham}}, \bibinfo {author} {\bibfnamefont
  {D.~J.}\ \bibnamefont {Twitchen}}, \bibinfo {author} {\bibfnamefont
  {R.}~\bibnamefont {Hanson}}, \ and\ \bibinfo {author} {\bibfnamefont {T.~H.}\
  \bibnamefont {Taminiau}},\ }\href {\doibase 10.1038/ncomms11526} {\bibfield
  {journal} {\bibinfo  {journal} {Nat. Commun.}\ }\textbf {\bibinfo {volume}
  {7}},\ \bibinfo {pages} {11526} (\bibinfo {year} {2016})}\BibitemShut
  {NoStop}%
\bibitem [{\citenamefont {Kalb}\ \emph {et~al.}(2016)\citenamefont {Kalb},
  \citenamefont {Cramer}, \citenamefont {Twitchen}, \citenamefont {Markham},
  \citenamefont {Hanson},\ and\ \citenamefont
  {Taminiau}}]{kalb_experimental_2016}%
  \BibitemOpen
  \bibfield  {author} {\bibinfo {author} {\bibfnamefont {N.}~\bibnamefont
  {Kalb}}, \bibinfo {author} {\bibfnamefont {J.}~\bibnamefont {Cramer}},
  \bibinfo {author} {\bibfnamefont {D.~J.}\ \bibnamefont {Twitchen}}, \bibinfo
  {author} {\bibfnamefont {M.}~\bibnamefont {Markham}}, \bibinfo {author}
  {\bibfnamefont {R.}~\bibnamefont {Hanson}}, \ and\ \bibinfo {author}
  {\bibfnamefont {T.~H.}\ \bibnamefont {Taminiau}},\ }\href {\doibase
  10.1038/ncomms13111} {\bibfield  {journal} {\bibinfo  {journal} {Nat.
  Commun.}\ }\textbf {\bibinfo {volume} {7}},\ \bibinfo {pages} {13111}
  (\bibinfo {year} {2016})}\BibitemShut {NoStop}%
\bibitem [{\citenamefont {Gaebler}\ \emph {et~al.}(2016)\citenamefont
  {Gaebler}, \citenamefont {Tan}, \citenamefont {Lin}, \citenamefont {Wan},
  \citenamefont {Bowler}, \citenamefont {Keith}, \citenamefont {Glancy},
  \citenamefont {Coakley}, \citenamefont {Knill}, \citenamefont {Leibfried},\
  and\ \citenamefont {Wineland}}]{gaebler_high-fidelity_2016}%
  \BibitemOpen
  \bibfield  {author} {\bibinfo {author} {\bibfnamefont {J.~P.}\ \bibnamefont
  {Gaebler}}, \bibinfo {author} {\bibfnamefont {T.~R.}\ \bibnamefont {Tan}},
  \bibinfo {author} {\bibfnamefont {Y.}~\bibnamefont {Lin}}, \bibinfo {author}
  {\bibfnamefont {Y.}~\bibnamefont {Wan}}, \bibinfo {author} {\bibfnamefont
  {R.}~\bibnamefont {Bowler}}, \bibinfo {author} {\bibfnamefont {A.~C.}\
  \bibnamefont {Keith}}, \bibinfo {author} {\bibfnamefont {S.}~\bibnamefont
  {Glancy}}, \bibinfo {author} {\bibfnamefont {K.}~\bibnamefont {Coakley}},
  \bibinfo {author} {\bibfnamefont {E.}~\bibnamefont {Knill}}, \bibinfo
  {author} {\bibfnamefont {D.}~\bibnamefont {Leibfried}}, \ and\ \bibinfo
  {author} {\bibfnamefont {D.~J.}\ \bibnamefont {Wineland}},\ }\href {\doibase
  10.1103/PhysRevLett.117.060505} {\bibfield  {journal} {\bibinfo  {journal}
  {Phys. Rev. Lett.}\ }\textbf {\bibinfo {volume} {117}},\ \bibinfo {pages}
  {060505} (\bibinfo {year} {2016})}\BibitemShut {NoStop}%
\bibitem [{\citenamefont {Harty}\ \emph {et~al.}(2014)\citenamefont {Harty},
  \citenamefont {Allcock}, \citenamefont {Ballance}, \citenamefont {Guidoni},
  \citenamefont {Janacek}, \citenamefont {Linke}, \citenamefont {Stacey},\ and\
  \citenamefont {Lucas}}]{harty_high-fidelity_2014}%
  \BibitemOpen
  \bibfield  {author} {\bibinfo {author} {\bibfnamefont {T.~P.}\ \bibnamefont
  {Harty}}, \bibinfo {author} {\bibfnamefont {D.~T.~C.}\ \bibnamefont
  {Allcock}}, \bibinfo {author} {\bibfnamefont {C.~J.}\ \bibnamefont
  {Ballance}}, \bibinfo {author} {\bibfnamefont {L.}~\bibnamefont {Guidoni}},
  \bibinfo {author} {\bibfnamefont {H.~A.}\ \bibnamefont {Janacek}}, \bibinfo
  {author} {\bibfnamefont {N.~M.}\ \bibnamefont {Linke}}, \bibinfo {author}
  {\bibfnamefont {D.~N.}\ \bibnamefont {Stacey}}, \ and\ \bibinfo {author}
  {\bibfnamefont {D.~M.}\ \bibnamefont {Lucas}},\ }\href {\doibase
  10.1103/PhysRevLett.113.220501} {\bibfield  {journal} {\bibinfo  {journal}
  {Phys. Rev. Lett.}\ }\textbf {\bibinfo {volume} {113}},\ \bibinfo {pages}
  {220501} (\bibinfo {year} {2014})}\BibitemShut {NoStop}%
\bibitem [{\citenamefont {Lange}\ \emph {et~al.}(2010)\citenamefont {Lange},
  \citenamefont {Wang}, \citenamefont {Rist{\`e}}, \citenamefont
  {Dobrovitski},\ and\ \citenamefont {Hanson}}]{lange_universal_2010}%
  \BibitemOpen
  \bibfield  {author} {\bibinfo {author} {\bibfnamefont {G.~d.}\ \bibnamefont
  {Lange}}, \bibinfo {author} {\bibfnamefont {Z.~H.}\ \bibnamefont {Wang}},
  \bibinfo {author} {\bibfnamefont {D.}~\bibnamefont {Rist{\`e}}}, \bibinfo
  {author} {\bibfnamefont {V.~V.}\ \bibnamefont {Dobrovitski}}, \ and\ \bibinfo
  {author} {\bibfnamefont {R.}~\bibnamefont {Hanson}},\ }\href {\doibase
  10.1126/science.1192739} {\bibfield  {journal} {\bibinfo  {journal}
  {Science}\ }\textbf {\bibinfo {volume} {330}},\ \bibinfo {pages} {60}
  (\bibinfo {year} {2010})}\BibitemShut {NoStop}%
\bibitem [{\citenamefont {Abobeih}\ \emph {et~al.}()\citenamefont {Abobeih},
  \citenamefont {Cramer}, \citenamefont {Bakker}, \citenamefont {Kalb},
  \citenamefont {Twitchen}, \citenamefont {Markham},\ and\ \citenamefont
  {Taminiau}}]{abobeih_one-second_2018}%
  \BibitemOpen
  \bibfield  {author} {\bibinfo {author} {\bibfnamefont {M.~H.}\ \bibnamefont
  {Abobeih}}, \bibinfo {author} {\bibfnamefont {J.}~\bibnamefont {Cramer}},
  \bibinfo {author} {\bibfnamefont {M.~A.}\ \bibnamefont {Bakker}}, \bibinfo
  {author} {\bibfnamefont {N.}~\bibnamefont {Kalb}}, \bibinfo {author}
  {\bibfnamefont {D.~J.}\ \bibnamefont {Twitchen}}, \bibinfo {author}
  {\bibfnamefont {M.}~\bibnamefont {Markham}}, \ and\ \bibinfo {author}
  {\bibfnamefont {T.~H.}\ \bibnamefont {Taminiau}},\ }\href
  {http://arxiv.org/abs/1801.01196} {\bibinfo  {journal} {arXiv:1801.01196}\
  }\BibitemShut {NoStop}%
\bibitem [{\citenamefont {Kolesov}\ \emph {et~al.}(2012)\citenamefont
  {Kolesov}, \citenamefont {Xia}, \citenamefont {Reuter}, \citenamefont
  {St{\"o}hr}, \citenamefont {Zappe}, \citenamefont {Meijer}, \citenamefont
  {Hemmer},\ and\ \citenamefont {Wrachtrup}}]{kolesov_optical_2012}%
  \BibitemOpen
\bibfield  {journal} {  }\bibfield  {author} {\bibinfo {author} {\bibfnamefont
  {R.}~\bibnamefont {Kolesov}}, \bibinfo {author} {\bibfnamefont
  {K.}~\bibnamefont {Xia}}, \bibinfo {author} {\bibfnamefont {R.}~\bibnamefont
  {Reuter}}, \bibinfo {author} {\bibfnamefont {R.}~\bibnamefont {St{\"o}hr}},
  \bibinfo {author} {\bibfnamefont {A.}~\bibnamefont {Zappe}}, \bibinfo
  {author} {\bibfnamefont {J.}~\bibnamefont {Meijer}}, \bibinfo {author}
  {\bibfnamefont {P.~R.}\ \bibnamefont {Hemmer}}, \ and\ \bibinfo {author}
  {\bibfnamefont {J.}~\bibnamefont {Wrachtrup}},\ }\href {\doibase
  10.1038/ncomms2034} {\bibfield  {journal} {\bibinfo  {journal} {Nat.
  Commun.}\ }\textbf {\bibinfo {volume} {3}},\ \bibinfo {pages} {1029}
  (\bibinfo {year} {2012})}\BibitemShut {NoStop}%
\bibitem [{\citenamefont {Christle}\ \emph {et~al.}(2015)\citenamefont
  {Christle}, \citenamefont {Falk}, \citenamefont {Andrich}, \citenamefont
  {Klimov}, \citenamefont {Hassan}, \citenamefont {Son}, \citenamefont
  {Janz{\'e}n}, \citenamefont {Ohshima},\ and\ \citenamefont
  {Awschalom}}]{christle_isolated_2015}%
  \BibitemOpen
  \bibfield  {author} {\bibinfo {author} {\bibfnamefont {D.~J.}\ \bibnamefont
  {Christle}}, \bibinfo {author} {\bibfnamefont {A.~L.}\ \bibnamefont {Falk}},
  \bibinfo {author} {\bibfnamefont {P.}~\bibnamefont {Andrich}}, \bibinfo
  {author} {\bibfnamefont {P.~V.}\ \bibnamefont {Klimov}}, \bibinfo {author}
  {\bibfnamefont {J.~U.}\ \bibnamefont {Hassan}}, \bibinfo {author}
  {\bibfnamefont {N.~T.}\ \bibnamefont {Son}}, \bibinfo {author} {\bibfnamefont
  {E.}~\bibnamefont {Janz{\'e}n}}, \bibinfo {author} {\bibfnamefont
  {T.}~\bibnamefont {Ohshima}}, \ and\ \bibinfo {author} {\bibfnamefont
  {D.~D.}\ \bibnamefont {Awschalom}},\ }\href {\doibase 10.1038/nmat4144}
  {\bibfield  {journal} {\bibinfo  {journal} {Nat. Mater.}\ }\textbf {\bibinfo
  {volume} {14}},\ \bibinfo {pages} {160} (\bibinfo {year} {2015})}\BibitemShut
  {NoStop}%
\bibitem [{\citenamefont {Sukachev}\ \emph {et~al.}(2017)\citenamefont
  {Sukachev}, \citenamefont {Sipahigil}, \citenamefont {Nguyen}, \citenamefont
  {Bhaskar}, \citenamefont {Evans}, \citenamefont {Jelezko},\ and\
  \citenamefont {Lukin}}]{sukachev_silicon-vacancy_2017}%
  \BibitemOpen
  \bibfield  {author} {\bibinfo {author} {\bibfnamefont {D.~D.}\ \bibnamefont
  {Sukachev}}, \bibinfo {author} {\bibfnamefont {A.}~\bibnamefont {Sipahigil}},
  \bibinfo {author} {\bibfnamefont {C.~T.}\ \bibnamefont {Nguyen}}, \bibinfo
  {author} {\bibfnamefont {M.~K.}\ \bibnamefont {Bhaskar}}, \bibinfo {author}
  {\bibfnamefont {R.~E.}\ \bibnamefont {Evans}}, \bibinfo {author}
  {\bibfnamefont {F.}~\bibnamefont {Jelezko}}, \ and\ \bibinfo {author}
  {\bibfnamefont {M.~D.}\ \bibnamefont {Lukin}},\ }\href {\doibase
  10.1103/PhysRevLett.119.223602} {\bibfield  {journal} {\bibinfo  {journal}
  {Phys. Rev. Lett.}\ }\textbf {\bibinfo {volume} {119}},\ \bibinfo {pages}
  {223602} (\bibinfo {year} {2017})}\BibitemShut {NoStop}%
\bibitem [{\citenamefont {Iwasaki}\ \emph {et~al.}()\citenamefont {Iwasaki},
  \citenamefont {Miyamoto}, \citenamefont {Taniguchi}, \citenamefont
  {Siyushev}, \citenamefont {Metsch}, \citenamefont {Jelezko},\ and\
  \citenamefont {Hatano}}]{iwasaki_tin-vacancy_2017}%
  \BibitemOpen
  \bibfield  {author} {\bibinfo {author} {\bibfnamefont {T.}~\bibnamefont
  {Iwasaki}}, \bibinfo {author} {\bibfnamefont {Y.}~\bibnamefont {Miyamoto}},
  \bibinfo {author} {\bibfnamefont {T.}~\bibnamefont {Taniguchi}}, \bibinfo
  {author} {\bibfnamefont {P.}~\bibnamefont {Siyushev}}, \bibinfo {author}
  {\bibfnamefont {M.~H.}\ \bibnamefont {Metsch}}, \bibinfo {author}
  {\bibfnamefont {F.}~\bibnamefont {Jelezko}}, \ and\ \bibinfo {author}
  {\bibfnamefont {M.}~\bibnamefont {Hatano}},\ }\href@noop {} {\bibinfo
  {journal} {arXiv:1708.03576}\ }\BibitemShut {NoStop}%
\bibitem [{\citenamefont {Morse}\ \emph {et~al.}(2017)\citenamefont {Morse},
  \citenamefont {Abraham}, \citenamefont {DeAbreu}, \citenamefont {Bowness},
  \citenamefont {Richards}, \citenamefont {Riemann}, \citenamefont {Abrosimov},
  \citenamefont {Becker}, \citenamefont {Pohl}, \citenamefont {Thewalt},\ and\
  \citenamefont {Simmons}}]{morse_photonic_2017}%
  \BibitemOpen
\bibfield  {journal} {  }\bibfield  {author} {\bibinfo {author} {\bibfnamefont
  {K.~J.}\ \bibnamefont {Morse}}, \bibinfo {author} {\bibfnamefont {R.~J.~S.}\
  \bibnamefont {Abraham}}, \bibinfo {author} {\bibfnamefont {A.}~\bibnamefont
  {DeAbreu}}, \bibinfo {author} {\bibfnamefont {C.}~\bibnamefont {Bowness}},
  \bibinfo {author} {\bibfnamefont {T.~S.}\ \bibnamefont {Richards}}, \bibinfo
  {author} {\bibfnamefont {H.}~\bibnamefont {Riemann}}, \bibinfo {author}
  {\bibfnamefont {N.~V.}\ \bibnamefont {Abrosimov}}, \bibinfo {author}
  {\bibfnamefont {P.}~\bibnamefont {Becker}}, \bibinfo {author} {\bibfnamefont
  {H.-J.}\ \bibnamefont {Pohl}}, \bibinfo {author} {\bibfnamefont {M.~L.~W.}\
  \bibnamefont {Thewalt}}, \ and\ \bibinfo {author} {\bibfnamefont
  {S.}~\bibnamefont {Simmons}},\ }\href {\doibase 10.1126/sciadv.1700930}
  {\bibfield  {journal} {\bibinfo  {journal} {Sci. Adv.}\ }\textbf {\bibinfo
  {volume} {3}},\ \bibinfo {pages} {e1700930} (\bibinfo {year}
  {2017})}\BibitemShut {NoStop}%
\bibitem [{\citenamefont {Lim}\ \emph {et~al.}()\citenamefont {Lim},
  \citenamefont {Welinski}, \citenamefont {Ferrier}, \citenamefont {Goldner},\
  and\ \citenamefont {Morton}}]{lim_coherent_2017}%
  \BibitemOpen
  \bibfield  {author} {\bibinfo {author} {\bibfnamefont {H.-J.}\ \bibnamefont
  {Lim}}, \bibinfo {author} {\bibfnamefont {S.}~\bibnamefont {Welinski}},
  \bibinfo {author} {\bibfnamefont {A.}~\bibnamefont {Ferrier}}, \bibinfo
  {author} {\bibfnamefont {P.}~\bibnamefont {Goldner}}, \ and\ \bibinfo
  {author} {\bibfnamefont {J.~J.~L.}\ \bibnamefont {Morton}},\ }\href
  {http://arxiv.org/abs/1712.00435} {\bibinfo  {journal} {arXiv:1712.00435}\
  }\BibitemShut {NoStop}%
\bibitem [{\citenamefont {Briegel}\ \emph {et~al.}(1998)\citenamefont
  {Briegel}, \citenamefont {D{\"u}r}, \citenamefont {Cirac},\ and\
  \citenamefont {Zoller}}]{briegel_quantum_1998}%
  \BibitemOpen
\bibfield  {journal} {  }\bibfield  {author} {\bibinfo {author} {\bibfnamefont
  {H.-J.}\ \bibnamefont {Briegel}}, \bibinfo {author} {\bibfnamefont
  {W.}~\bibnamefont {D{\"u}r}}, \bibinfo {author} {\bibfnamefont {J.~I.}\
  \bibnamefont {Cirac}}, \ and\ \bibinfo {author} {\bibfnamefont
  {P.}~\bibnamefont {Zoller}},\ }\href {\doibase 10.1103/PhysRevLett.81.5932}
  {\bibfield  {journal} {\bibinfo  {journal} {Phys. Rev. Lett.}\ }\textbf
  {\bibinfo {volume} {81}},\ \bibinfo {pages} {5932} (\bibinfo {year}
  {1998})}\BibitemShut {NoStop}%
\bibitem [{\citenamefont {Rozp{\k e}dek}\ \emph {et~al.}()\citenamefont
  {Rozp{\k e}dek}, \citenamefont {Goodenough}, \citenamefont {Ribeiro},
  \citenamefont {Kalb}, \citenamefont {Vivoli}, \citenamefont {Reiserer},
  \citenamefont {Hanson}, \citenamefont {Wehner},\ and\ \citenamefont
  {Elkouss}}]{rozpedek_realistic_2017}%
  \BibitemOpen
  \bibfield  {author} {\bibinfo {author} {\bibfnamefont {F.}~\bibnamefont
  {Rozp{\k e}dek}}, \bibinfo {author} {\bibfnamefont {K.}~\bibnamefont
  {Goodenough}}, \bibinfo {author} {\bibfnamefont {J.}~\bibnamefont {Ribeiro}},
  \bibinfo {author} {\bibfnamefont {N.}~\bibnamefont {Kalb}}, \bibinfo {author}
  {\bibfnamefont {V.~C.}\ \bibnamefont {Vivoli}}, \bibinfo {author}
  {\bibfnamefont {A.}~\bibnamefont {Reiserer}}, \bibinfo {author}
  {\bibfnamefont {R.}~\bibnamefont {Hanson}}, \bibinfo {author} {\bibfnamefont
  {S.}~\bibnamefont {Wehner}}, \ and\ \bibinfo {author} {\bibfnamefont
  {D.}~\bibnamefont {Elkouss}},\ }\href@noop {} {\bibinfo  {journal}
  {arXiv:1705.00043}\ }\BibitemShut {NoStop}%
\bibitem [{\citenamefont {Nickerson}\ \emph {et~al.}(2014)\citenamefont
  {Nickerson}, \citenamefont {Fitzsimons},\ and\ \citenamefont
  {Benjamin}}]{nickerson_freely_2014}%
  \BibitemOpen
\bibfield  {journal} {  }\bibfield  {author} {\bibinfo {author} {\bibfnamefont
  {N.~H.}\ \bibnamefont {Nickerson}}, \bibinfo {author} {\bibfnamefont {J.~F.}\
  \bibnamefont {Fitzsimons}}, \ and\ \bibinfo {author} {\bibfnamefont {S.~C.}\
  \bibnamefont {Benjamin}},\ }\href {\doibase 10.1103/PhysRevX.4.041041}
  {\bibfield  {journal} {\bibinfo  {journal} {Phys. Rev. X}\ }\textbf {\bibinfo
  {volume} {4}},\ \bibinfo {pages} {041041} (\bibinfo {year}
  {2014})}\BibitemShut {NoStop}%
\end{thebibliography}%

\end{document}